\documentclass[]{aa} 
\usepackage{comment}
\usepackage{natbib}
\usepackage{graphicx}
\usepackage{footmisc}
\usepackage{float}
\usepackage{caption}
\usepackage[caption=false]{subfig}
\usepackage[colorlinks=true,citecolor=blue]{hyperref}
\usepackage{multirow}
\usepackage{amsmath}
\usepackage{nccmath}
\usepackage{color}
\usepackage{array} 
\usepackage{ulem}
\usepackage[flushleft]{threeparttable}
\DeclareTextFontCommand{\textroman}{\fontlibertine}
\usepackage{txfonts}

\begin{document}

\title{Is there \ion{Na}{i} in the atmosphere of HD~209458b?}

\subtitle{Effect of the centre-to-limb variation and Rossiter-McLaughlin effect in transmission spectroscopy studies}

   \author{N.~Casasayas-Barris\inst{1,2} \and E.~Pallé\inst{1,2} \and F.~Yan\inst{3} \and G.~Chen\inst{4} \and R.~Luque\inst{1,2} \and M.~Stangret\inst{1,2} \and E.~Nagel\inst{5} \and M.~Zechmeister\inst{3} \and  M.~Oshagh\inst{3} \and  J.~Sanz-Forcada\inst{6} \and L.~Nortmann\inst{1,2} \and F.~J.~Alonso-Floriano\inst{7} \and P.~J.~Amado\inst{8} \and J.~A.~Caballero\inst{6} \and S.~Czesla\inst{5} \and S.~Khalafinejad\inst{14} \and M.~López-Puertas\inst{8} \and J.~López-Santiago\inst{9,10} \and K.~Molaverdikhani\inst{11} \and D. Montes\inst{15} \and A.~Quirrenbach\inst{14} \and A.~Reiners\inst{3} \and I.~Ribas\inst{12,13} \and A.~Sánchez-López\inst{8} \and M.~R.~Zapatero~Osorio\inst{6}
 }


   \institute{\label{inst:iac}Instituto de Astrofísica de Canarias, Vía Láctea s/n, 38205 La Laguna, Tenerife, Spain
              \\
              \email{nuriacb@iac.es}
         \and
             \label{inst:ull}Departamento de Astrofísica, Universidad de La Laguna, Spain
        \and
            \label{inst:gott}Institut für Astrophysik, Georg-August-Universität, Friedrich- Hund-Platz 1, 37077 Göttingen, Germany
        \and
            \label{inst:pmo}Key Laboratory of Planetary Sciences, Purple Mountain Observatory, Chinese Academy of Sciences, Nanjing 210033, China
        \and
            \label{inst:ham}Hamburger Sternwarte, Universität Hamburg, Gojenbergsweg 112, 21029 Hamburg, Germany
        \and
            \label{inst:calb}Centro de Astrobiología (CSIC-INTA), ESAC, Camino bajo del castillo s/n, 28692 Villanueva de la Cañada, Madrid, Spain
        \and
            \label{inst:leiden}Leiden Observatory, Leiden University, Postbus 9513, 2300 RA, Leiden, The Netherlands
        \and
            \label{inst:iaa}Instituto de Astrofísica de Andalucía (IAA-CSIC), Glorieta de la Astronomía s/n, 18008 Granada, Spain
        \and 
            \label{inst:uciii}Department of Signal Theory and Communications, Universidad Carlos III de Madrid, Av. de la Universidad 30, Leganés, 28911 Madrid, Spain
        \and 
            \label{inst:ucm}Gregorio Mara\~non Health Research Institute, Doctor Esquerdo 46, 28007, Madrid, Spain
        \and
            \label{inst:mpia}Max-Planck-Institut für Astronomie, Königstuhl 17, 69117 Heidelberg, Germany
        \and
            \label{inst:ice}Institut de Ciències de l'Espai (ICE,CSIC), Campus UAB, c/ de Can Magrans s/n, 08193 Bellaterra, Barcelona, Spain
        \and 
            \label{inst:ieec}Institut d'Estudis Espacials de Catalunya (IEEC), 08034 Barcelona, Spain
        \and
            \label{inst:uheidel}Landessternwarte, Zentrum für Astronomie der Universität Heidelberg, Königstuhl 12, 69117 Heidelberg, Germany
        \and
            \label{inst:ucm}Departamento de Física de la Tierra y Astrofísica and IPARCOS-UCM (Intituto de Física de Partículas y del Cosmos de la UCM), Facultad de Ciencias Físicas, Universidad Complutense de Madrid, 28040, Madrid, Spain
        }


   \date{Received 29 November 2019 / Accepted 19 February 2020}

  \abstract{HD~209458b was the first transiting planet discovered, and the first for which an atmosphere, in particular \ion{Na}{i}, was detected. With time, it has become one of the most frequently studied planets, with a large diversity of atmospheric studies using low- and high-resolution spectroscopy. Here, we present transit spectroscopy observations of HD~209458b using the HARPS-N and CARMENES spectrographs. We fit the Rossiter-McLaughlin effect by combining radial velocity data from both instruments (nine transits in total), measuring a projected spin-orbit angle of $-1.6\pm0.3~{\rm deg}$. We also present the analysis of high-resolution transmission spectroscopy around the \ion{Na}{i} region at $590~{\rm nm}$, using a total of five transit observations. In contrast to previous studies where atmospheric \ion{Na}{i} absorption is detected, we find that for all of the nights, whether individually or combined, the transmission spectra can be explained by the combination of the centre-to-limb variation and the Rossiter-McLaughlin effect. This is also observed in the time-evolution maps and transmission light curves, but at lower signal-to-noise ratio. Other strong lines such as H$\alpha$, \ion{Ca}{ii} IRT, the \ion{Mg}{i} triplet region, and \ion{K}{i} D1 are analysed, and are also consistent with the modelled effects, without considering any contribution from the exoplanet atmosphere.
   Thus, the transmission spectrum reveals no detectable \ion{Na}{i} absorption in HD~209458b. We discuss how previous pioneering studies of this benchmark object may have overlooked these effects. While for some star-planet systems these effects are small, for other planetary atmospheres the results reported in the literature may require revision. 
 }


   \maketitle
%
\section{Introduction}
\label{intro}

HD~209458b was the first exoplanet discovered that transits in front of its host star \citep{Charbonneau2000,Henry2000ApJ...529L..41H}, and it was also the first exoplanet for which an atmosphere was detected \citep{2002ApJ...568..377C}. Charbonneau and collaborators detected neutral sodium (\ion{Na}{i}) in HD~209458b using  data from the Space Telescope Imaging Spectrograph (STIS) onboard the {\it Hubble Space Telescope} ({\it HST}). The same data were used by  \citet{Sing2008HD209ApJ...686..658S}, who resolved both the \ion{Na}{i} D2 and D1 lines. With high-dispersion spectroscopy, \ion{Na}{i} absorption was also detected by \citet{2008SnellenHD209} using the High Dispersion Spectrograph (HDS) at the Subaru telescope, and it was tentatively confirmed by \citet{jensen2011} using the Hobby–Eberly Telescope. Additionally, \citet{Simon2009IAUS..253..520A} detected \ion{Na}{i} in two UVES (Ultraviolet and Visual Echelle Spectrograph) data sets taken with the Very Large Telescope (VLT). 

The Rossiter-McLaughlin (RM) effect \citep{ros24,mcl24} is produced when a planet transits in front of its host star, occulting a part of the stellar disc. When the stellar rotation is assumed to be in the same direction as the orbital motion of the planet, the most strongly blue-shifted wings of the stellar disc are partly occulted when the transit starts, yielding red-shifted lines; similarly, the end of transit yields blue-shifted lines. The opposite situation (stellar spin opposite to the revolution of the planet) yields red-shifted lines at the beginning of the transit, and blue-shifted lines at the end. Thus the RM effect allows us to know the geometry of the stellar rotation as compared to the planetary motion, and even to calculate the projected spin-orbit angle. This effect is identified in the radial velocity curve. If this effect is not correctly taken into account, it may lead to a misidentification of the planetary atmospheric spectral lines during transits, which are only a manifestation of the RM effect.

Centre-to-limb variations (CLVs) may also potentially affect the line profiles during transits. The stellar continuum in the photosphere has a lower intensity near the stellar limb than at the centre of the disc. This effect is related to the optical depth of the photosphere. A more subtle effect arises from Fraunhofer lines that form at different heights in the stellar atmosphere; the balance between the lines that form at different heights depends on the limb angle and stellar latitude \citep{abe35,app67}. The strength of the CLV-induced effect can be of the same order as signals found from hot-Jupiter atmospheres \citep{Yan2017A&A...603A..73Y,StefanCLV2015A&A...582A..51C, Khalafinehad2017A&A...598A.131K}.

Here, we report our observations of the benchmark planet HD~209458b using the high-dispersion spectrographs HARPS-N and CARMENES. With the combination of several transits with each instrument and the use of models for the CLV and RM effects, we revisited the observational evidence for the detection of \ion{Na}{i} in the atmosphere of HD~209458b.

This paper is organised as follows. In Section~\ref{obs} we detail the observations. In Section~\ref{ssec:RMfit} we estimate the obliquity of the HD~209458b system. The methods for extracting the high-resolution transmission spectra and light curves are explained in Section~\ref{method}. In Section~\ref{sec:results} we present the results obtained in the analysis of high-resolution transmission spectra around the \ion{Na}{i} and other lines, and the analysis of systematic effects. In Section~\ref{sec:comp} our results are compared with previous studies of the same planet around \ion{Na}{i}. The discussion and conclusions are presented in Section~\ref{sec:disc}.

\section{Observations}
\label{obs}
We used archival transit observations of HD 209458b obtained with the HARPS-N and CARMENES spectrographs. A total of nine transits are available in the archives. However, only five of them are considered in this atmospheric analysis. The other four nights are discarded because the signal-to-noise ratio (S/N) of the observations was low. The information related to the observations is summarised in Table~\ref{Tab:Obs}. Nights 3 and 7 are the same night, observed simultaneously with the HARPS-N and CARMENES spectrographs.

\begin{table*}[]
\centering
\caption{Observing log of the HD~209458b transit observations.}
\resizebox{\textwidth}{!}{\begin{tabular}{cllccccccl}
\hline\hline
Night & Tel. & Instrument & Date of     &   $t_\mathrm{exp}$ & $N_\mathrm{obs}$ & S/N\tablefootmark{a} & S/N\tablefootmark{b} & Analysis & Fibre B \\
      &      & & observation & [s] &               & \ion{Na}{i} order & \ion{Na}{i} core & atmosphere/RM fit& \\ \hline
\\[-1em]
1 & TNG & HARPS-N & 2015-09-26 & 600 & 32 & 162  & 51&Yes/Yes & \ion{Na}{i} sky emission\\ 
\\[-1em]
2 & TNG & HARPS-N & 2016-07-25 &   600 & 42 & 63 & 20&No/Yes & \ion{Na}{i} sky emission\\ 
\\[-1em]
3 & TNG & HARPS-N & 2016-09-16 &   600 & 45 & 136 & 43&Yes/Yes & Fabry-Pérot observations \\
\\[-1em]
4 & TNG & HARPS-N & 2017-07-16 &   300/600 & 38 & 64 & 20&No/Yes & \ion{Na}{i} sky emission \\ 
\\[-1em]
5 & TNG & HARPS-N & 2017-09-07 & 300 & 61 &  108 & 33&Yes/Yes & Fabry-Pérot observations\\
\\[-1em]
6 & CA 3.5~m & CARMENES & 2016-09-09 &  180 & 38 &  82& 37 &No/Yes & No observable \ion{Na}{i} sky emission\\ 
\\[-1em]
7 & CA 3.5~m & CARMENES & 2016-09-16 &  180 & 70 & 87  & 40&Yes/Yes & \ion{Na}{i} sky emission \\ 
\\[-1em]
8 & CA 3.5~m & CARMENES & 2016-11-08 &  180 & 80 & 54  & 22&No/Yes & \ion{Na}{i} sky emission\\ 
\\[-1em]
9 & CA 3.5~m & CARMENES & 2018-09-05 & 192 & 82 & 84  & 38&Yes/Yes & No observable \ion{Na}{i} sky emission \\ 

\\[-1em]
\hline
\end{tabular}}\\
\tablefoot{\tablefoottext{a}{Averaged S/N per extracted pixel calculated in the \ion{Na}{i} order (53 for HARPS-N and 104 for CARMENES) for each night.} \tablefoottext{b}{Averaged S/N in the \ion{Na}{i} D2 and D1 line cores (calculated in $\pm5$~km\,s$^{-1}$ centred on the cores). This calculation is performed by dividing the flux of each pixel by its photon noise.}}
\label{Tab:Obs}
\end{table*}

\subsection{HARPS-N observations}
A total of five transit observations of HD~209458b are publicly available in the Telescopio Nazionale Galileo (TNG) archive. The HARPS-N (High Accuracy Radial velocity Planet Searcher for the Northern hemisphere) spectrograph (\citealt{HARPSN12003Msngr.114...20M}, \citealt{HARPSN22012SPIE.8446E..1VC}) is mounted on the $3.58$m TNG telescope, located at the Observatorio del Roque de los Muchachos (ORM, La Palma), and covers the optical range from $383$ nm to $690$ nm. The observations were carried out under programs A32TAC\_41 and A35TAC\_14. These observations were performed with continuous exposures during the transit, and some additional exposures were taken before and after. 

The two observations with lower S/N ($\sim 60$) in the \ion{Na}{i} order (nights 2 and 4 in Table~\ref{Tab:Obs}) were discarded for the atmospheric analysis. For night 1 no data were taken during the ingress, and only three stellar spectra are observed before the transit. The sky emission is observed in fibre B in nights 1, 2, and 4. In nights 3 and 5, fibre B was used for Fabry-Pérot observations, which means that we lack information on possible sky emission during the night. The position of the telluric \ion{Na}{i} emission lines for these nights (if existing) is expected to be around $8$~km\,s$^{-1}$ and $12$~km\,s$^{-1}$ from the stellar \ion{Na}{i} line cores, respectively. At these separations, the presence of sky contamination would affect the results. The sky emission of night 1 was corrected for by subtracting the sky spectrum in fibre B from the science spectrum in fibre A. Unfortunately, the sensitivity of fibre A and B might not be the same. In this case, it would result in an imperfect removal of the telluric sky emission, which  would be propagated throughout the process. As this emission is observed inside the \ion{Na}{i} line wings and not in the continuum, we are not able to see it in the stellar spectrum (fibre A), and consequently, the sensitivity of the two fibres cannot be compared. However, the residuals related to an imperfect sky emission correction would not reproduce the results presented here. If existing, however, they could affect the amplitude of the results. No interstellar \ion{Na}{i} is observable in the spectra.

\subsection{CARMENES observations}

Four more transits were observed with the CARMENES spectrograph \citep[Calar Alto high-Resolution search for M dwarfs with Exo-earths with Near-infrared and optical Echelle Spectrographs;][]{CARMENES, CARMENES18} located on the Calar Alto Observatory. CARMENES is a two-channel spectrograph that simultaneously covers the optical range from $520$ to $960$ nm and the near-infrared range from $960$ to $1710$ nm. In this study we only use the optical observations. The observations were carried out under programs H16-3.5-24, H16-3.5-22, and H18-3.5-22, and the infrared data were studied by \citet{SanchezLopez2019} and \citet{AlonsoFloriano2019HD209}. The observing strategy was the same as for the HARPS-N observations. 

We discarded night 8 observations from our atmospheric study because of their lower S/N (around $50$). The observations of night 6 could not be used because a technical problem of the telescope led to the loss of the first half of the transit. For this reason, only nights 7 and 9 were used in the atmospheric analysis. In all CARMENES observations, fibre B was used to monitor the sky contribution. When the sky spectra were checked, telluric \ion{Na}{i} sky emission was observed in nights 7 and 8.

\section{Estimating the obliquity of HD~209458b}
\label{ssec:RMfit}

The radial velocity anomaly due to the RM effect can be observed in the stellar radial velocity measurements of HD~209458 in all transit observations. In order to obtain the system parameters related to this effect, we performed a joint analysis of the radial velocity values from HARPS-N and CARMENES. To this end, we used all five nights observed with HARPS-N and all four nights observed with CARMENES. Radial velocities for both instruments were obtained using the SERVAL programme \citep{SERVAL}, which uses least-squares fitting with a high S/N template created by co-adding all available spectra of the star to compute the radial velocities.

The fitting procedure of the RM effect model to the radial velocity data was performed using the Markov chain Monte Carlo (MCMC) algorithm implemented in {\tt emcee} \citep{emcee2013PASP..125..306F}. We used the RM effect model presented in \citet{2005Ohta} together with a circular orbital radial velocity (both contributions implemented in {\tt PyAstronomy} \citep{PyAstronomy2019ascl.soft06010C} as {\tt modelSuite.RmcL} and {\tt modelSuite.radVel}, respectively). The RM model depends on the orbital period ($P$), the transit epoch ($T_c$), ratio of the planet-to-star radius ($R_p/R_{\star}$), the angular rotation velocity of the host star ($\Omega$), the linear limb-darkening coefficient ($\epsilon$), the inclination of the orbit ($i$), the inclination of the stellar rotation axis ($i_{\star}$), the sky-projected angle between the stellar rotation axis and the normal to the plane of planetary orbit ($\lambda$), and the scaled semi-major axis ($a/R_{\star}$). The circular orbit radial velocity contribution depends on $P$, $T_c$, the stellar velocity semi-amplitude ($K_{\star}$), and the offset with respect to the null radial velocity ($\Delta v$).

As in \citet{2017CasasayasB}, we fixed $i_{\star}$ to $90~\mathrm{deg}$, and $P$, $a/R_{\star}$, $R_p/R_{\star}$, $R_{\star}$, and $i$ were taken from the values presented in Table~\ref{tab:Param}, while the other parameters remained free. Because we fitted different HARPS-N and CARMENES nights, we needed to take into account that some free parameters should be fitted jointly for all nights and instruments, and others that could change. $\Omega$, $\lambda$, and $T_c$ were jointly fitted in all cases, while for each night we considered different $\Delta v$ and $K_{\star}$ values. We note that the offset between the model and the data can vary from night to night because in addition to the systematic velocity, the radial velocity information contains possible instrumental and stellar activity effects that are reflected as additional offsets to the data. $K_{\star}$ could also be affected by activity and become different for different nights \citep{MahmoudActivity2018A&A...619A.150O}. Finally, because CARMENES and HARPS-N cover different wavelength regions, we defined two different linear limb-darkening coefficient parameters, $\epsilon^{\rm C}$ and $\epsilon^{\rm H}$ for the CARMENES and HARPS-N data, respectively.

\begin{table*}[]
\small
\centering
\caption{Physical and orbital parameters of the HD~209458 system from previous studies}.
\begin{tabular}{lll}
\\ \hline \hline
\\[-1em]
 Description  & Symbol [Units] & Value \\ \hline
 \\[-1em]
 \multicolumn{3}{c}{\dotfill\it Stellar parameters\dotfill}\\\noalign{\smallskip}
   \\[-1em]
\quad  Effective temperature\tablefootmark{a} & $T_{\rm eff}$ [K]& $6065\pm50$\\
  \\[-1em]
\quad  Projected rotation speed\tablefootmark{b} & $v\sin i_{\star}$ [km\,s$^{-1}$]& $4.70\pm0.16$\\
\quad Surface gravity\tablefootmark{a} & $\log g$ [cgs]& $4.361^{+0.007}_{-0.008}$\\
\quad  Metallicity\tablefootmark{a} & [Fe/H] & $0.00\pm0.05$\\
\quad Stellar mass\tablefootmark{a} & $M_{\star}$ [$\rm{M_{\odot}}$]& $1.119\pm0.033$\\
 \\[-1em]
\quad  Stellar radius\tablefootmark{a} &$R_{\star}$ [$\rm{R_{\odot}}$]& $1.155^{+0.014}_{-0.016}$\\ 
 \\[-1em]
 \multicolumn{3}{c}{\dotfill\it Planet parameters\dotfill}\\\noalign{\smallskip}
   \\[-1em]
                
 \quad Planet mass\tablefootmark{a} & $M_{\rm p}$ [$\rm{M_{Jup}}$]&  $0.682^{+0.015}_{-0.014}$\\
 \\[-1em]
 \quad Planet radius\tablefootmark{a} & $R_{\rm p}$ [$\rm{R_{Jup}}$]& $1.359^{+0.016}_{-0.019}$ \\
 \\[-1em]
 \quad Equilibrium temperature\tablefootmark{a} &$T_{\rm eq}$ [K]& $1449\pm12$\\ 
 \\[-1em]
 \multicolumn{3}{c}{\dotfill\it Transit parameters\dotfill}\\\noalign{\smallskip}
   \\[-1em]
 \quad Epoch\tablefootmark{c} & $T_{\rm c}$ [BJD$_{\rm TDB}$] & $2454560.80588\pm 0.00008$\\
  \\[-1em]
 \quad Period\tablefootmark{d} & $P$ [day] & $3.52474859\pm0.00000038$\\
  \\[-1em]
 \quad Transit duration\tablefootmark{e} & $T_{14}$ [h]& $2.978\pm0.051$\\
  \\[-1em]
 \quad Full in-transit duration\tablefootmark{e} & $T_{23}$ [h]& $2.254\pm0.058$\\
 \\[-1em]
 \multicolumn{3}{c}{\dotfill\it System parameters\dotfill}\\\noalign{\smallskip}
   \\[-1em]
 \quad Sem-imajor axis\tablefootmark{a} &$a$ [au]& $0.04707^{+0.00046}_{-0.00047}$\\
 \\[-1em]
  \quad Scaled semi-major axis\tablefootmark{a} &$a/R_{\star}$& $8.76\pm0.04$\\
 \\[-1em]
 \quad Inclination\tablefootmark{a} & $i$ [deg]& $86.71\pm0.05$\\
   \\[-1em]
 \quad Systemic velocity\tablefootmark{f} &$\gamma$ [km\,s$^{-1}$]& $-14.741\pm0.002$\\
  \\[-1em]
\quad Stellar velocity semi-amplitude\tablefootmark{d} &$K_{\star}$ [m\,s$^{-1}$]& $84.27^{+0.69}_{-0.70}$\\
  \\[-1em]
 \quad Projected obliquity\tablefootmark{b} &$\lambda$ [deg]& $-4.4\pm1.4$\\\lasthline
\end{tabular}
\tablefoot{\tablefoottext{a}{\citet{Torres2008ApJ...677.1324T}.} \tablefoottext{b}{\citet{Winn2005HD209}.} \tablefoottext{c}{\citet{Evans2015MNRAS.451..680E}.} \tablefoottext{d}{\citet{Bonomo2017A&A...602A.107B}.} \tablefoottext{e}{\citet{HD209Richardson_2006}.} \tablefoottext{f}{\citet{HD2092004Naef}.} 
}
\label{tab:Param}
\end{table*}

\begin{figure}[]
\centering
\includegraphics[width=0.49\textwidth]{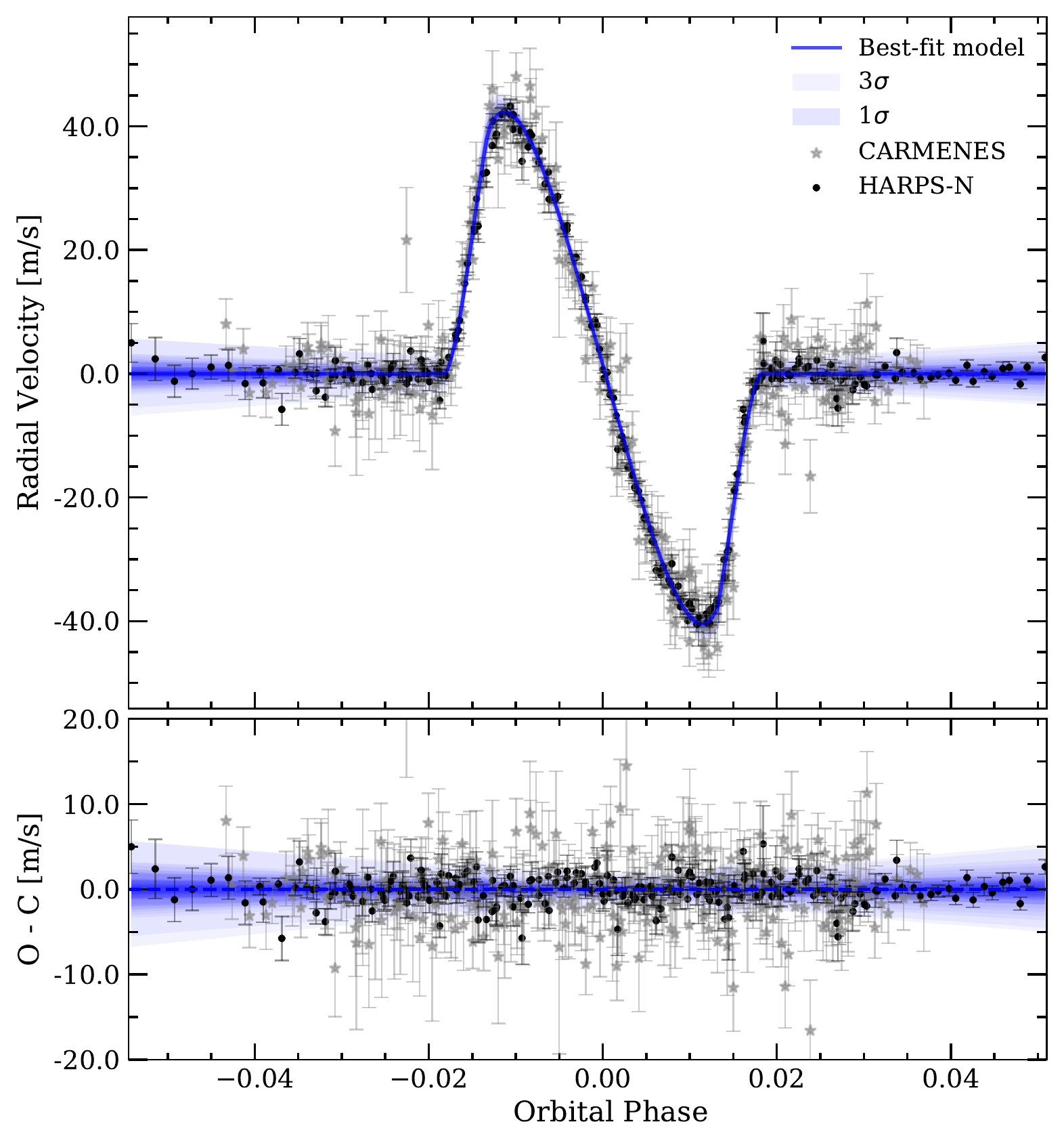}
\caption{Radial velocity anomaly due to the RM effect (top panel) and residuals between the data and model (bottom panel). The detrended stellar radial velocity values observed with CARMENES (four transits) are presented as grey stars. HARPS-N data sets (five transits) are presented as black dots. The blue line shows the model resulting from the combination of the HARPS-N and CARMENES best-fit models. In light blue we present the $1\sigma$ and $3\sigma$ uncertainties of the model.}
\label{fig:RM_all}
\end{figure}

The system was analysed using $100$ walkers with $10^5$ steps. The first $7~000$ steps were discarded as burn-in. Each step was started at a random point near the expected values from the literature, and $\lambda$ was constrained to $\pm180~{\rm deg}$. On the other hand, $\Omega$ was constrained to (0.25,0.50)~rad\,d$^{-1}$. The median values of the posteriors were adopted as the best-fit values, and their error bars correspond to the $1\sigma$ statistical errors at the corresponding percentiles. The MCMC results are presented in Table~\ref{tab:RM_res} (Case 1) and the detrended data and best-fit model in Figure~\ref{fig:RM_all}. The radial velocity curves were detrended using the best-fit $K_{\star}$ and $\Delta v$ values of each night. The data and best-fit model for each night and correlation diagrams for the probability distribution are shown in Figures~\ref{fig:RM_indiv} and \ref{fig:RM_corner} of the appendix, respectively. All $\Delta v$ and $K_{\star}$ best-fit values are presented in Table~\ref{tab:RM_res_add}.

\renewcommand{\thefootnote}{\fnsymbol{footnote}}
\begin{table}[]
\centering
\caption{RM effect MCMC best-fit values with $1\sigma$ and $3\sigma$ uncertainties using different assumptions.}
\begin{tabular}{lcccc}
\hline\hline
\\[-1em]
Symbol & Units & Value & $1\sigma$ & $3\sigma$\\ \hline
\\[-1em]
\\[-1em]
T$_c$ & BJD &  $2452826.62980$ & $\pm0.00009$ & $^{+0.00026}_{-0.00025}$\\ 
\\[-1em]
$\epsilon^{\rm H}$ & ... & $0.94$& $\pm0.01$ & $\pm0.03$\\ 
\\[-1em]
$\epsilon^{\rm C}$  & ... &  $0.84$ & $\pm0.03$ & $\pm 0.08$\\
\\[-1em]
 \multicolumn{5}{c}{\dotfill\it Case 1 \dotfill}\\\noalign{\smallskip}
\\[-1em]
$\lambda$ & deg & $-1.6$ & $\pm0.3$ & $\pm0.8$\\ 
\\[-1em]
$\Omega$ & rad\,d$^{-1}$ & $0.540$ & $\pm0.002$ & $\pm0.006$ \\ 
$i_{\star}$  & $\deg$ &  $90$ (fixed)& ...& ...\\
\\[-1em]
$e$  & ... &  $0$ (fixed)& ... & ...\\
\\[-1em]
$\omega$  & $\deg$ &  $90$ (fixed)&... & ...\\
\\[-1em]
 \multicolumn{5}{c}{\dotfill\it Case 2 \dotfill}\\\noalign{\smallskip}
\\[-1em]
$\lambda$ & deg & $-1.6$ & $\pm0.3$ & $\pm0.8$\\ 
\\[-1em]
$\Omega$ & rad\,d$^{-1}$ & $0.63$ & $\pm0.04$ & $^{+0.06}_{-0.08}$ \\ 
\\[-1em]
$i_{\star}$  & $\deg$ &  $58.7$ & $^{+7.9}_{-5.2}$ & $^{+19.3}_{-7.5}$\\
\\[-1em]
$e$  & ... &  $0$ (fixed)& ... & ...\\
\\[-1em]
$\omega$  & $\deg$ &  $90$ (fixed)&... & ...\\
\\[-1em]
 \multicolumn{5}{c}{\dotfill\it Case 3 \dotfill}\\\noalign{\smallskip}
\\[-1em]
$\lambda$ & deg & $-1.5$ & $^{+0.4}_{-0.3}$ & $\pm1.0$\\ 
\\[-1em]
$\Omega$ & rad\,d$^{-1}$ & $0.540$ & $\pm0.002$ & $^{+0.007}_{-0.005}$ \\ 
\\[-1em]
$i_{\star}$  & $\deg$ &  $90$ (fixed)& ...& ...\\
\\[-1em]
$e$  & ... &  $0.05$ & $^{+0.05}_{-0.04}$ & $^{+0.09}_{-0.05}$\\
\\[-1em]
$\omega$  & $\deg$ &  $89.3$ & $^{+4.7}_{-5.7}$ & $^{+13.9}_{-14.4}$\\
\\[-1em]
\\[-1em]
\hline
\end{tabular}
\label{tab:RM_res}
\end{table}
\renewcommand{\thefootnote}{\arabic{footnote}}

The best-fit $T_c$ value can be compared to the value measured by \citet{Evans2015MNRAS.451..680E} by propagating over different orbits using the orbital period from Table~\ref{tab:Param}. We obtain a transit centre of $2454560.8061\pm0.0002~{\rm BJD}$, which corresponds to a difference of ${\sim 19}~{\rm s}$ with the measurement reported by \citet{Evans2015MNRAS.451..680E}. Both measurements are consistent considering the reported uncertainties. The error bar of the propagated transit centre was calculated by considering the $T_c$ uncertainty from Table~\ref{tab:RM_res} and the orbital period uncertainty from Table~\ref{tab:Param}.

In Table~\ref{tab:RM_res_add} we observe that $K_{\star}$ has different values for different nights. When these values are compared with the value reported by \citet{Bonomo2017A&A...602A.107B} (see Table~\ref{tab:Param}), nights 5, 7, and 9 do not present consistent results (at $3\sigma$). In addition to the possible $K_{\star}$ variations induced by stellar activity \citep{MahmoudActivity2018A&A...619A.150O}, telluric contamination is particularly high when radial velocities are extracted. During the extraction with the SERVAL pipeline, we realised that telluric lines are not entirely masked during the process and introduce radial velocity gradients, which leads to $K_{\star}$ variations. Thus, the differences observed in different nights might be produced by the combination of these two factors.

With one transit of HD~209458b, \citet{Winn2005HD209} measured an almost aligned system with $\lambda=-4.4\pm1.4$ and $v\sin i_{\star} = 4.70\pm0.16$~km\,s$^{-1}$. On the other hand, \citet{Albrecht2012ApJ...757...18A} found $\lambda = -5\pm7~{\rm deg}$ and $v\sin i_{\star} = 4.4\pm0.2$~km\,s$^{-1}$. Here, with nine transits, we measure a spin-orbit angle of $-1.6\pm0.3~{\rm deg}$ and an angular rotation velocity of $0.540\pm0.002~$rad\,d$^{-1}$. We note that the slight difference (smaller than $2\sigma$) between the previous results and our $\lambda$ estimate might arise because $\lambda$ and $\Omega$ are degenerate \citep{Brown2017MNRAS.464..810B,Albrecht2012ApJ...757...18A}. In addition, as reported by \citet{Bourrier_2017}, a traditional velocimetric analysis of the RM effect could lead to biases in the measured spin-orbit angle as a result of changes in the local CCF shape.

In order to determine the dependence of the reported results on the $i_{\star}$ value, we fitted the data by leaving this parameter free, constrained to (0--180)\,deg, hereafter Case~2. As expected, we observe a strong correlation between $\Omega$ and $i_{\star}$, while the remaining parameters maintain consistency with the results obtained assuming $i_{\star}=90~\deg$ (i.e. Case~1). This same exercise was performed with the eccentricity (now fixing $i_{\star}$ to $90$\,deg), hereafter Case~3. We left the eccentricity ($e$) and the argument of periapsis passage ($\omega$) as free parameters, constrained to $(0,1)$ and (0--180)\,deg, respectively. We observe that the radial velocity offset $\Delta v$ and $K_{\star}$ parameters are strongly correlated with $\omega$ and $e$, respectively. On the other hand, even the best-fit parameters are consistent with the values obtained under the assumption of a circular orbit (i.e. Cases~1 and~2), the projected obliquity $\lambda$ presents a small correlation with $\omega$ value. The best-fit values obtained in these two cases are shown in Tables~\ref{tab:RM_res} and~\ref{tab:RM_res_add} as Cases~2 and~3, respectively. In both cases, the MCMC analysis was performed using $80$ walkers and $10^4$ steps.

\section{Methods}
\label{method}

\subsection{Transmission spectrum and light-curve extraction}
\label{sec:TS_TLC}
The HARPS-N observations were reduced with the HARPS-N Data reduction Software (DRS), version $3.7$ (\citealt{HARPSNDRS12014SPIE.9147E..8CC}, \citealt{HARPSNDRS22014ASPC..485..435S}). The DRS extracts the spectra order by order, and they are then flat-fielded. A blaze correction and the wavelength calibration are applied to each spectral order, and finally, all the spectral orders from each two-dimensional échelle spectrum are combined and resampled with a wavelength step of $0.01~\mathrm{\AA}$ into a one-dimensional spectrum. The spectra are referenced to the barycentric rest frame and the wavelengths are given in air.

CARMENES observations were processed with the CARMENES pipeline CARACAL \citep[CARMENES Reduction And Calibration; ][]{CARACAL}, which considers bias, flat-relative optimal extraction \citep{flatrelative}, cosmic-ray correction, and the wavelength calibration described in \citet{carmcalibration}. The reduced spectra are referenced to the terrestrial rest frame and the wavelengths are given in vacuum.

The transmission spectrum of each night was extracted as presented in \citet{Casasayas2018} and \citet{Casasayas2019}. In summary, we first corrected the telluric absorption contamination using Molecfit (\citealt{Molecfit1} and \citealt{Molecfit2}). Then, the spectra were shifted to the stellar rest frame using the stellar radial velocity semi-amplitude $K_{\star}=84.27$~m\,s$^{-1}$ measured by \citet{Bonomo2017A&A...602A.107B}, the barycentric radial velocity information, and the system velocity (see the physical and orbital parameters used in Table~\ref{tab:Param}). The RM radial-velocity anomaly during the transit was not considered when we moved the spectra to the stellar rest frame. After the stellar spectra were aligned, we combined all the data taken when the planet was not transiting to a high S/N master spectrum (master out-of-transit spectrum). After this, the ratio of all spectra by the master spectrum was computed, and these residual spectra were moved to the planet rest frame. For this, we computed the planet radial velocity semi-amplitude, $K_p=144.89$~km\,s$^{-1}$, using the parameters in Table~\ref{tab:Param}. Finally, the in-transit residuals between second and third contacts of the transit were combined to determine the individual transmission spectrum of each night. It should be noted that these operations were all carried out on continuum-normalised spectra. Therefore the transit is not visible outside the spectral lines.

A small difference with respect to the previous studies is that here, this master out-of-transit spectrum is computed using the S/N of the \ion{Na}{i} order as weights. The reason of using the weighted mean to compute this spectrum is that the S/N inside the \ion{Na}{i} line cores is low. In order to determine one transmission spectrum per instrument, we averaged the results of the individual nights using the mean S/N of each night as weights. 

As presented in \citet{YanKELT9} and \citet{Casasayas2019}, we measured the transmission light curves after computing the ratio of the spectra by the master-out spectrum, moved to the planet rest frame. These light curves were measured by integrating the flux (using trapezoidal integration) inside two different bandwidths: $0.4~{\rm \AA}$ and $0.75~{\rm \AA}$. We strongly note that computing the light curves as detailed here (after computing the ratio of spectra) produces different results than if we had followed the method presented in \citet{2008SnellenHD209} and \citet{Simon2009IAUS..253..520A}, for example. In these studies, the flux inside a passband centred on the stellar lines is averaged (using the stellar spectra) and is then compared with adjacent passbands of the same size. For this reason, the results cannot be directly compared. The light curves of different nights and instruments were combined by sorting the values measured at different time stamps in chronological order with respect to the centre of the transit. In this way, we avoided an interpolation to a common time axis.

In both transmission spectra and light curve analysis we used the transit centre obtained in the RM fitting presented in Section~\ref{ssec:RMfit} (see Table~\ref{tab:RM_res}). For the remaining parameters we assumed the literature values presented in Table~\ref{tab:Param}.

\subsection{Modelling the RM and CLV effects}
\label{sec:mod}

To evaluate the stellar variation during the transit, we modelled the CLV and RM effects in the \ion{Na}{i} lines as presented in recent studies such as \citet{YanKELT9}, \citet{Yan2019arXiv191100380Y}, \citet{Casasayas2019}, and \citet{StefanCLV2015A&A...582A..51C}. The stellar spectra were modelled using MARCS \citep{MARCS2008A&A...486..951G}, assuming solar abundance, local thermodynamic equilibrium (LTE), and the stellar parameters presented in Table~\ref{tab:Param}. With the spectroscopy made easy (SME) tool \citep{SMENew2017A&A...597A..16P} we were then able to compute the stellar spectra for different limb-darkening angles and instrumental resolutions. We used the line lists from the VALD database \citep{VALD3}. After this, the CLV for different orbital phases of the planet was modelled following \citet{Yan2017A&A...603A..73Y}, together with the RM effect as presented by \citet{Yan2019arXiv191100380Y}, assuming the system parameters obtained in Section~\ref{ssec:RMfit}. The differences observed in the modelled spectra when the obliquity ($\lambda$) was assumed to be zero, the value measured in this work or the measurements from \citet{Winn2005HD209} and \citet{Albrecht2012ApJ...757...18A} are not significant when compared with the data.  

\begin{figure}[]
\centering
\includegraphics[width=0.5\textwidth]{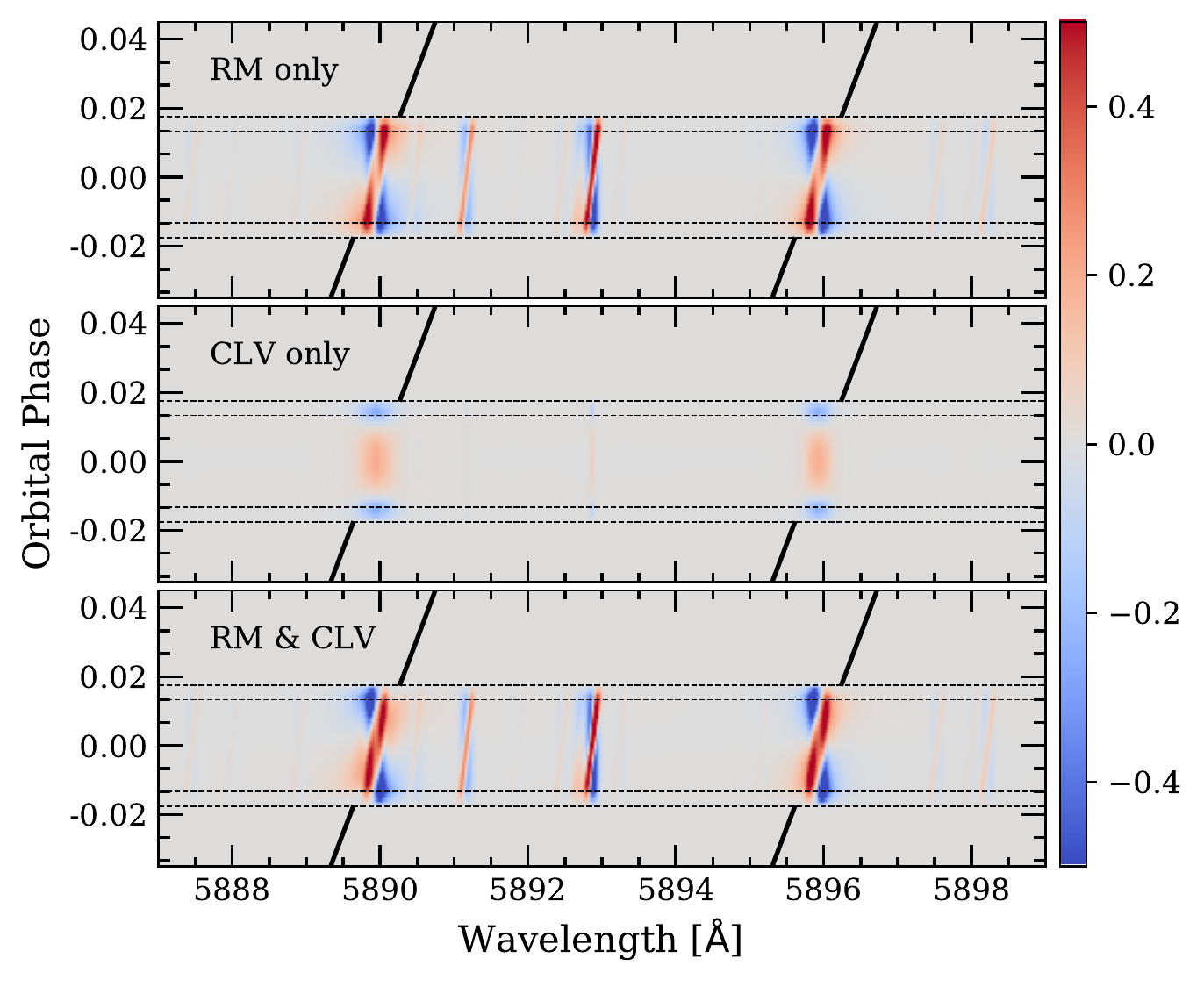}
\caption{Two-dimensional map of the modelled CLV and RM effects of HD~209458 system around the \ion{Na}{i} lines after dividing all stellar models by the out-of-transit spectrum, shown in the stellar rest frame. \textit{Top panel}: Model containing only the RM effect. \textit{Middle panel}: Model containing only the CLV effect. \textit{Bottom panel}: Model containing both the RM and CLV effects. The black solid lines show the calculated position of the planetary trail extrapolated to the out-of-transit time for better visualisation of the in-transit residuals. The horizontal dashed lines show the four contacts of the transit.} The colour bar describes the relative flux ($F_{\rm in} / F_{\rm out} - 1$) in $\%$. 
\label{fig:model_star}
\end{figure}

To observe the variation of the modelled stellar line profiles during the transit of HD~209458b, we divided each continuum-normalised stellar spectrum by the modelled out-of-transit spectrum, as we did for the data (see Section~\ref{sec:TS_TLC}). The evolution of these effects with the orbital phase of the planet can be observed in Figure~\ref{fig:model_star} in the form of what, hereafter, we call 2D maps. In these maps, we show the wavelength on the horizontal axis, the orbital phase of the planet on the vertical axis, and the relative flux is shown in colour. In this figure, we show the contribution of each individual effect and their combination, which clearly shows that the main contribution comes from the RM effect.

\begin{figure}[]
\centering
\includegraphics[width=0.5\textwidth]{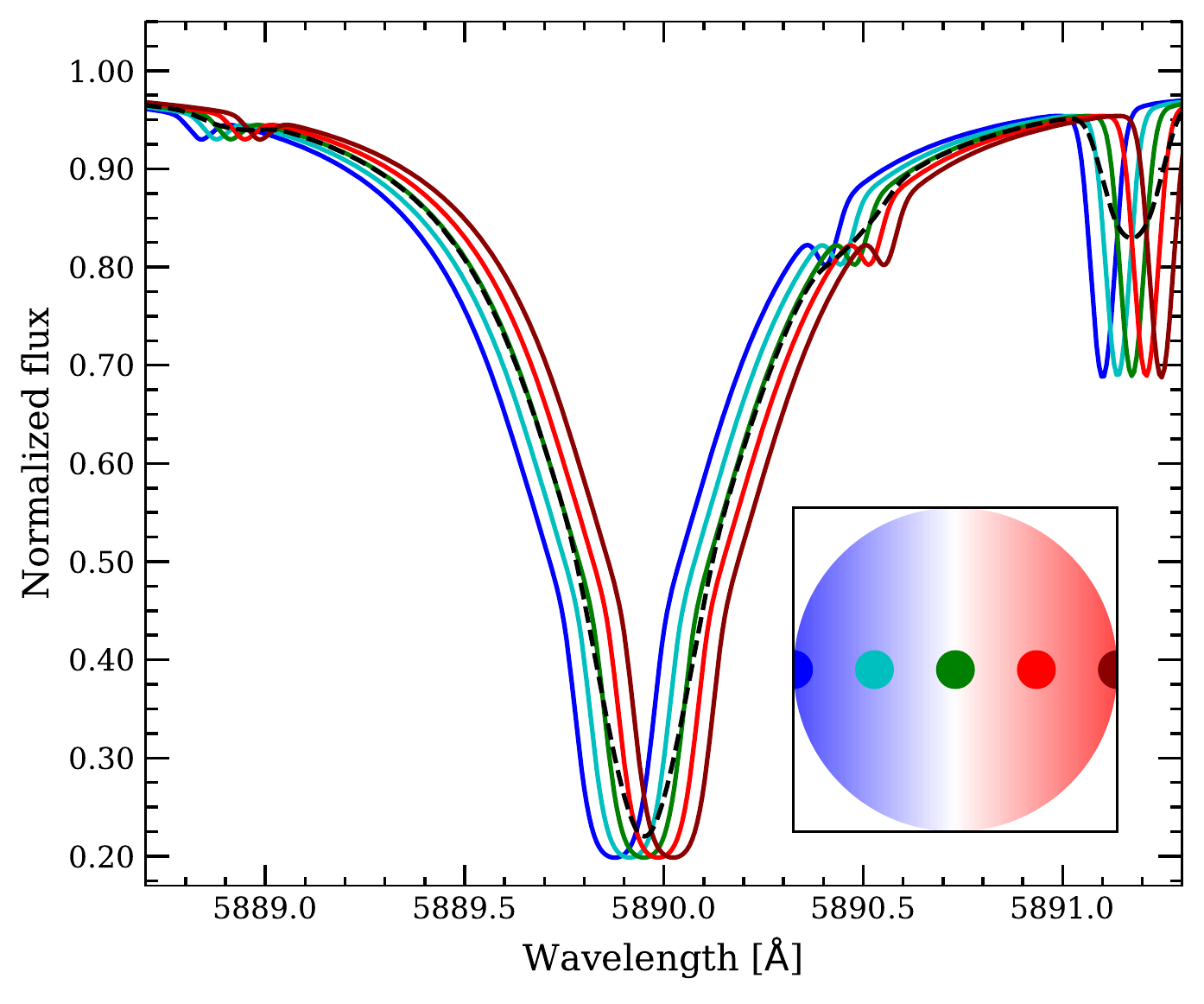}
\caption{Modelled stellar spectra around the \ion{Na}{i} D2 line of HD~209458 system, containing only the RM effect. The black dashed line shows the out-of-transit spectrum (integrated stellar disc). The coloured lines show the spectra of the regions that are blocked by the planet at five different orbital phases, which are then subtracted from the integrated stellar disc spectrum to compute the final stellar models. We note that these spectra are all normalised by their own continuum level for a better comparison of the line profile. However, in the real subtraction the blocked spectra represent only a small part of the light received from the stellar disc. The approximate position of the planet in each calculation (colours) is indicated in the subplot located in the bottom right corner of the main figure.}
\label{fig:RM_expl}
\end{figure}

The double feature caused by the RM effect (blue and red regions in the 2D maps) is easily recognised. This behaviour can be understood by studying Figure~\ref{fig:RM_expl}. When the planet crosses the stellar disc, it blocks the stellar light from different regions of the disc, which have different radial velocities. In our calculation, we first computed the integrated stellar disc spectrum when the planet was not transiting (out-of-transit). Then, we computed the spectrum of the regions of the disc that were blocked by the planet at different orbital phases. At a given position of the planet, the spectrum of the blocked region does not contribute to the final integrated disc spectrum, for this reason, it was then subtracted from the integrated stellar spectrum. In Figure~\ref{fig:RM_expl} we show the modelled out-of-transit spectrum profile (black) and the spectra of the regions that are blocked by the planet (colours) at different orbital phases (in this case, we only include the RM effect). As expected, the out-of-transit profile is centred at the laboratory position because it includes all velocities from the stellar disc. On the other hand, the coloured spectra are shifted with respect to the out-of-transit spectrum because they come from regions of the stellar disc that are described by different radial velocities. These different shifts and line shapes of the blocked spectra with respect to the integrated disc spectrum produce the double feature observed in the modelled RM effect. We note that in this figure, the spectra have been normalised by their continuum level for better visualisation of the line profile. The real contribution of the blocked regions to the out-of-transit spectrum is, of course, very small compared to the integrated disc spectrum, which produces the effects depicted in Figure~\ref{fig:model_star}.

\begin{figure}[]
\centering
\includegraphics[width=0.5\textwidth]{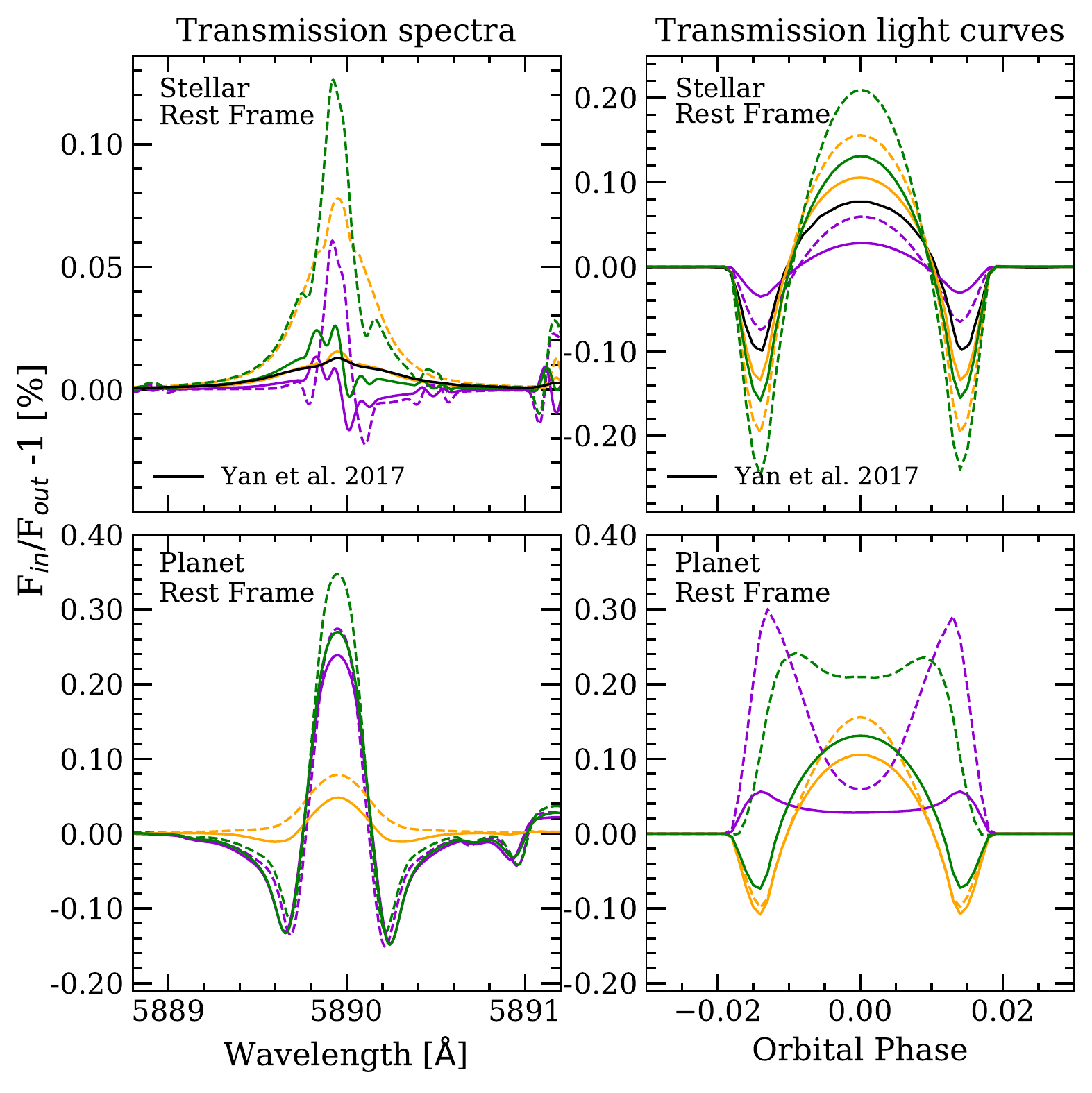}
\caption{Modelled transmission spectra (\textit{left}) and light curves (\textit{right}) around the \ion{Na}{i} D2 line of the system, containing only the CLV (yellow), only the RM effect (violet) and both effects together (green). In the \textit{top} row we show the results computed in the stellar rest frame, while in the \textit{bottom} row they are computed in the planet rest frame. For the modelled transmission spectra, we show with solid lines the results obtained by combining the data between the first and fourth transit contacts, while the dashed lines show only the data between the second and third contacts. For the transmission light curves, the solid lines correspond to a bandwidth of $0.75~{\rm \AA,}$ and the dashed lines correspond to $0.4~{\rm \AA}$ passband. The (black) line shown in the stellar rest frame results corresponds to the models presented in \citet{Yan2017A&A...603A..73Y}, which considered non-LTE effects in the stellar models calculation and measured the light curves using a different method.}
\label{fig:mod_TS_TLC}
\end{figure}

When the transmission spectrum and light curve models are computed, it is very important to follow the same method as is applied to the data. The CLV and RM effects in the transmission spectrum and light curves strongly depend on how we perform this calculation. For example, the transmission spectrum that was computed including only the spectra between the second and third transit contacts is different from the spectrum in which the ingress and egress spectra are included, especially when the modelled spectrum only contains the CLV effect (see its dependence on orbital phase in Figure~\ref{fig:model_star}). The CLV and RM effects are also partially compensated for when the in-transit exposures were moved to the stellar rest frame considering the RM radial velocity anomaly because of the misalignment that it introduces with respect to the out-of-transit spectra. On the other hand, for the transmission light curves, when we follow the method presented in \citet{2008SnellenHD209} and \citet{Simon2009IAUS..253..520A}, where the flux is measured in the \ion{Na}{i} stellar lines core (in the stellar rest frame), the curves are different than when we use the method presented here (in the planet rest frame and after computing the ratio between the individual spectra and the master-out spectrum; see also Section \ref{sec:disc}). This is particularly important for small bandwidths because the positive part (in relative flux) of the RM effect follows the radial velocities of the planet (see Figure~\ref{fig:model_star}). In Figure~\ref{fig:mod_TS_TLC} we show some transmission spectra and light-curve models that were computed using different methods. As an example, we also show the CLV effect model presented by \citet{Yan2017A&A...603A..73Y}, which is computed using non-LTE and in the stellar rest frame. We note how significant the effects become (especially the RM effect) in both transmission spectra and light curves when the spectra are moved to the planetary rest frame. In this particular case, for example, the CLV contribution in the transmission spectrum is more than four times smaller than the RM contribution. The importance of considering the RM effect for atmospheric studies was noted for the first time by \citet{Louden2015ApJ...814L..24L}. In this same paper, the RM effect in the transmission spectrum of HD~189733b was shown in the planet and in the stellar rest frames, noting how these effects are compensated for in the stellar rest frame.

\section{Analysis and results}
\label{sec:results}

\subsection{\ion{Na}{i} transmission spectrum}

\begin{figure}[]
\centering
\includegraphics[width=0.48\textwidth]{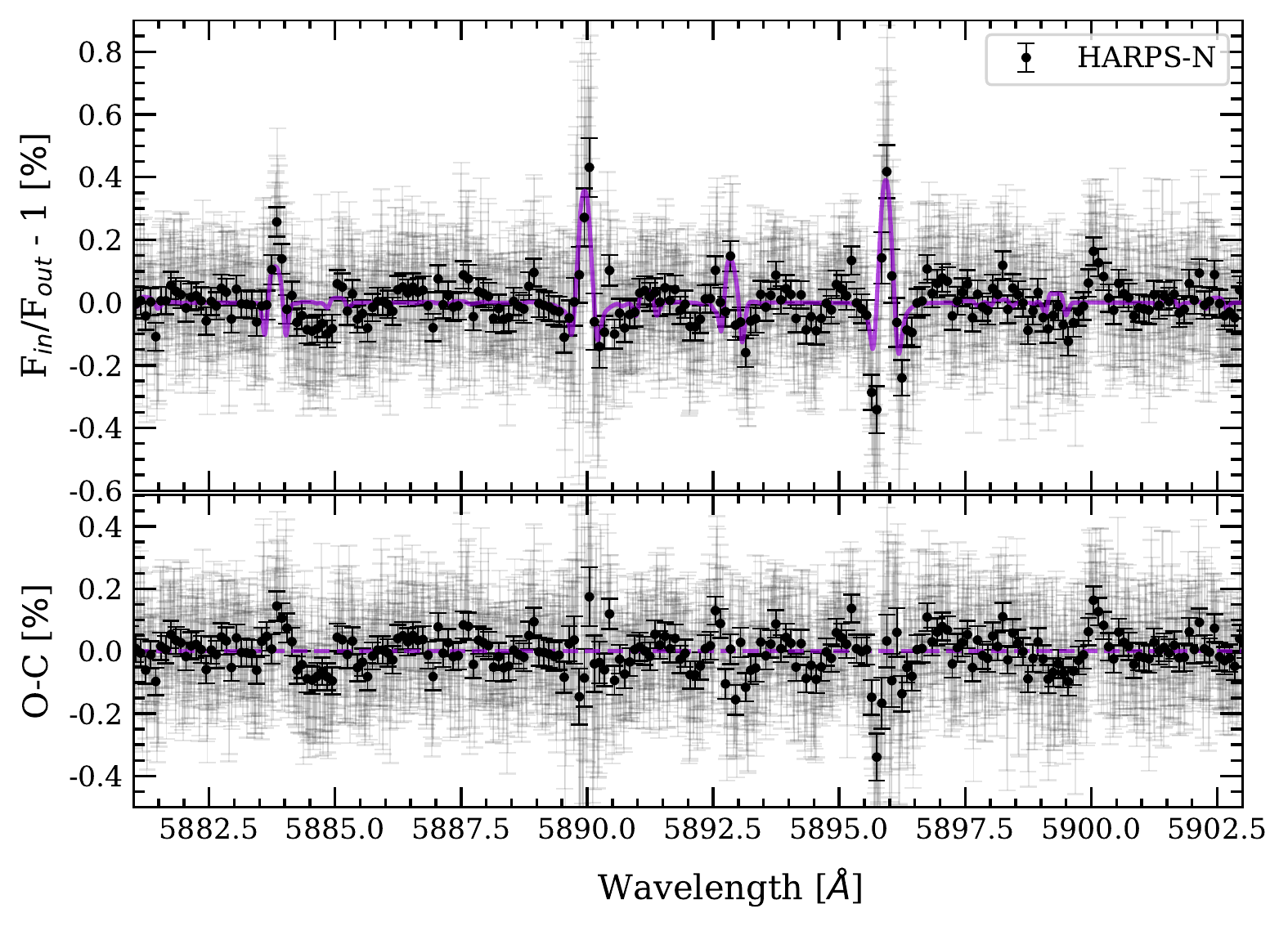}
\includegraphics[width=0.48\textwidth]{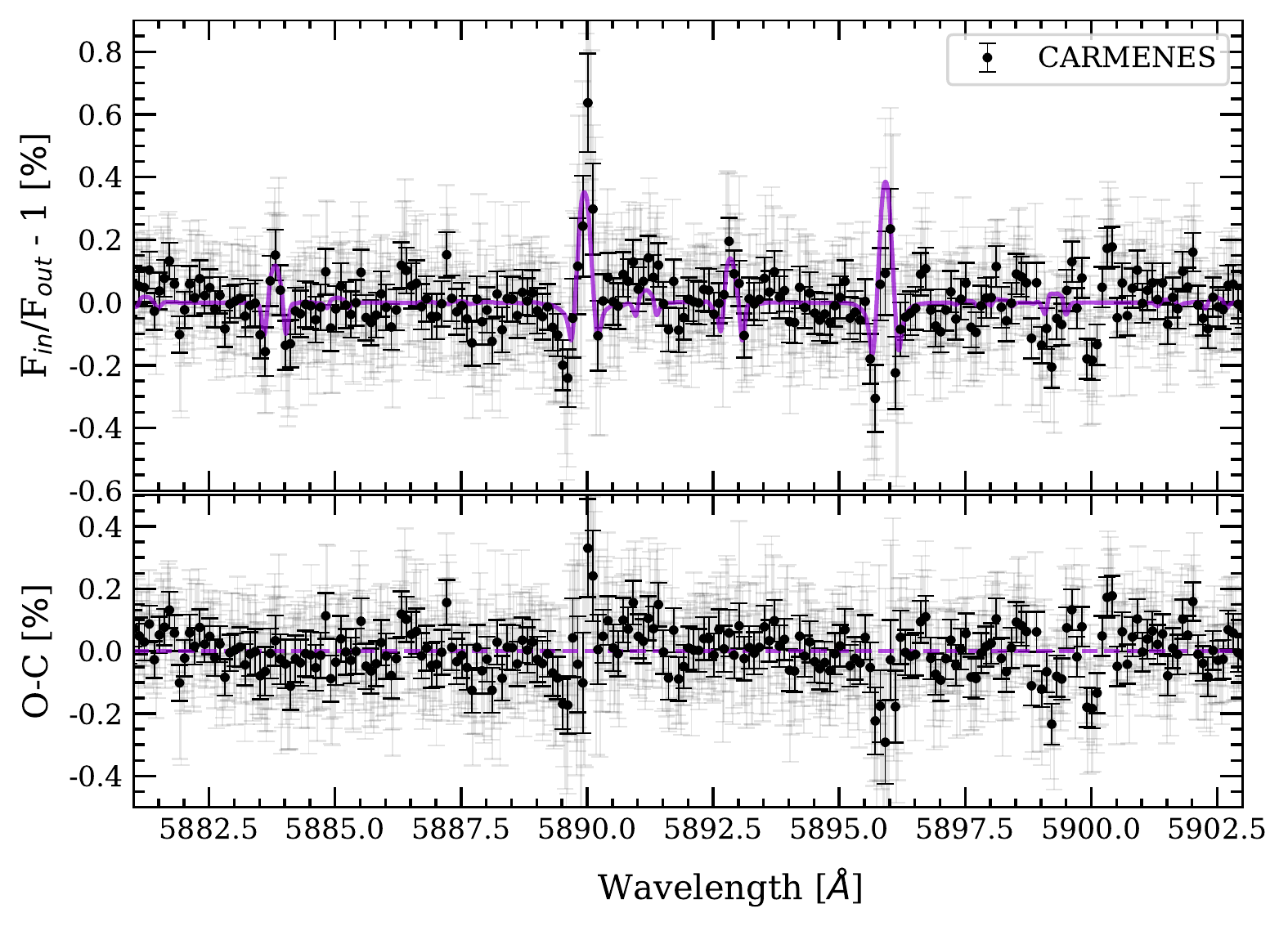}
\caption{HD~209458b transmission spectrum around the \ion{Na}{i} doublet after combining three nights observed with HARPS-N (\textit{top panel}) and two with CARMENES (\textit{bottom panel}). The light grey dots show the original data and the black dots the binned data using a bin width of $0.1~\mathrm{\AA}$. The purple line is the RM+CLV model. In each panel, the lower plot shows the residuals after the model is subtracted from the data.}
\label{fig:TS_comb}
\end{figure}

\begin{figure*}[]
\centering
\includegraphics[width=1\textwidth]{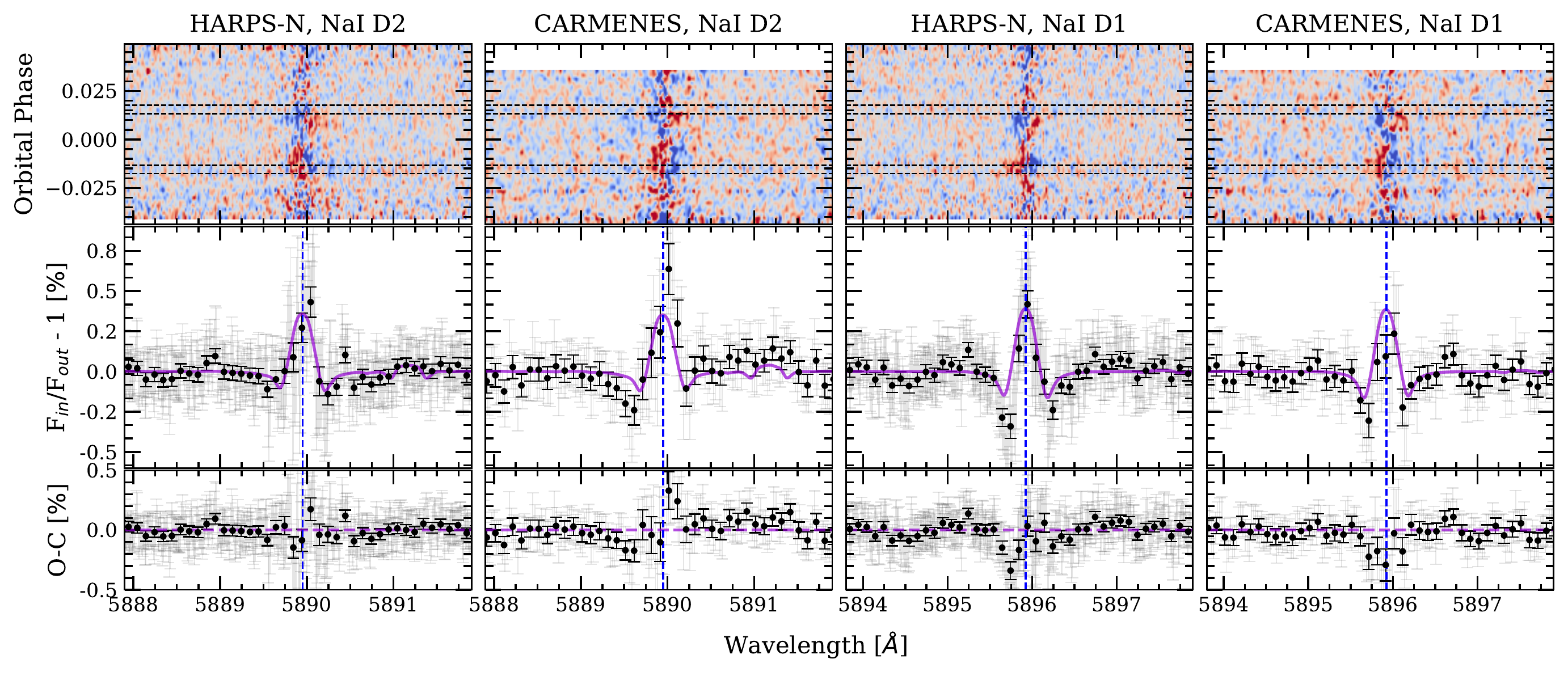}
\caption{Same transmission spectra as shown in Figure~\ref{fig:TS_comb}, but zooming into each line of the \ion{Na}{i} doublet. The blue vertical dashed lines show the laboratory value of the \ion{Na}{i} D2 or D1 line. In addition to the transmission spectrum of each line, we show the observed 2D maps around each particular line in the stellar rest frame. In these maps the spectra have been binned in orbital phase with a bin size of $0.002$. In colour we show the relative flux (blue and red correspond to negative and positive flux, respectively). The horizontal black dashed lines mark the four transit contacts.}
\label{fig:TS_comb_zoom}
\end{figure*}

The transmission spectra around the \ion{Na}{i} lines of HD~209458b are presented in Figure~\ref{fig:TS_comb} and a zoom-in on the lines in Figure~\ref{fig:TS_comb_zoom}. The results are the combination of three transits observed with HARPS-N and two with CARMENES. The individual transmission spectra obtained for each night are shown in Figures~\ref{fig:indiv_TS_HARPS} and \ref{fig:indiv_TS_CARM} in the Appendix. The laboratory wavelength of the \ion{Na}{i} D2 and D1 lines used here are $5889.951~{\rm \AA}$ and $5895.924~{\rm \AA}$ (in air), from the NIST database \citep{NIST_ASD}.

For all individual nights and instruments (including the noisiest nights, which are not shown here and were excluded from the analysis), instead of atmospheric absorption from the planet, we observe a pseudo-emission signal centred on the \ion{Na}{i} D2 and D1 line positions. When the data are compared with the model containing CLV and RM contributions (see Figure~\ref{fig:TS_comb_zoom}), we find that the data and the models are consistent. We point out that we directly compared the modelled effects with the data, without any fitting or re-scaling procedure. 

Other stellar lines in this wavelength range (at $5884$ and $5893~{\rm \AA}$ for example, consistent with \ion{Fe}{i} and \ion{Ni}{i}) also follow the expected modelled effects. We measure an averaged relative flux of $0.20\pm0.05~\%$ and $0.12\pm0.04~\%$ in the \ion{Na}{i} D2 and D1 lines, respectively, using a passband of $0.4~{\rm \AA}$ in the transmission spectrum obtained using HARPS-N data. For CARMENES, we get $0.31\pm0.08~\%$ and $0.07\pm0.06~\%$, respectively. If this same measurement is performed in our modelled transmission spectrum, which contains no planetary atmosphere, we obtain $0.22~\%$ in both lines. When subtracting the model from the transmission spectrum, some absorption-like residuals remain. This probably results from the combination of the limited model accuracy and the smaller S/N in the line cores.

The 2D maps around the \ion{Na}{i} doublet combining the different nights are presented in Figure~\ref{fig:TS_comb_zoom}. The S/N achieved around the \ion{Na}{i} line cores is relatively low (see values in Table~\ref{Tab:Obs}). Even with this low S/N, we are able to visually distinguish the behaviour observed in the modelled CLV+RM effects from Figure~\ref{fig:model_star} in the in-transit time region when the different nights are combined, however.
We note that we did not consider any correction related to the different exposure times of the different nights when the individual results were combined. For CARMENES, a similar exposure time was used in both nights, while for HARPS-N one of the nights has a very different exposure time ($300$\,s less than the others). This produces different smoothing of the signals because in $300$\,s the projected planetary radial velocity changes by around $0.9$~km\,s$^{-1}$ during the transit, although $0.9$~km\,s$^{-1}$ in HARPS-N correspond to around one pixel, that is, the effect is expected to be small.

\subsection{\ion{Na}{i} transmission light curves}

The combined transmission light curves of the \ion{Na}{i} D2 and D1 lines from both instruments are presented in Figure~\ref{fig:TLCNaID21}. In Appendix~\ref{ap:TLC} we present the individual transmission light curves of each instrument and line.

\begin{figure}[]
\centering
\includegraphics[width=0.4\textwidth]{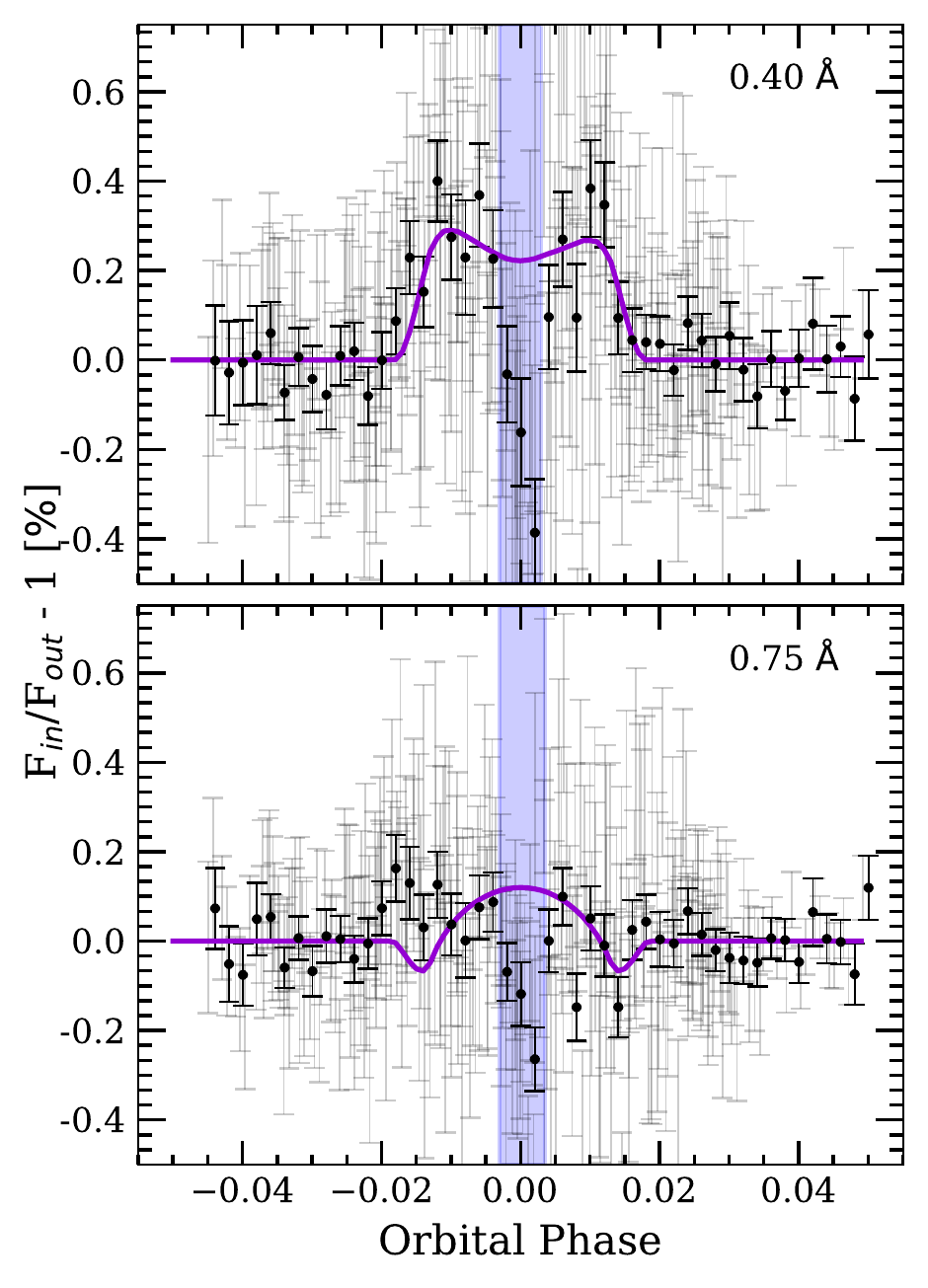}
\caption{\ion{Na}{i} observed transmission light curves calculated using the passbands of $0.40~{\rm \AA}$ (top panel) and $0.75~{\rm \AA}$ (bottom panel). In light grey we show the original data containing the combined HARPS-N and CARMENES results. The black dots are the data binned by $0.002$ in orbital phase. In purple we present the transmission light curve modelled considering the CLV and RM effects. In light blue we mark the region around zero orbital phase where the data present an unexpected behaviour.}
\label{fig:TLCNaID21}
\end{figure}

Because we measured the transmission light curves in the planet rest frame, the modelled light curves were computed in the same way. For small passbands (smaller than $0.4~{\rm \AA}$) we mainly included the positive (in relative flux) contribution of the RM effect in the light curves, while for larger passbands the negative contribution was also included, which decreased the overall effect (see Fig.~\ref{fig:model_star}) because they cancel each other out. This is the case when a passband of $0.75~{\rm \AA}$ is used, for example.

Both instruments and individual lines show similar results. For the small $0.4~{\rm \AA}$ passband, the light curves do not show a transit-like shape, but follow the model describing the RM effect. For broader passbands, the CLV+RM model curve shows very small change in amplitude, whic is difficult to observe at the S/N of our data, and indeed, the light curves observed in a $0.75~{\rm \AA}$ passband are mainly flat, with large scatter. In terms of the absorption depth, we measure $0.25~\%$ and $0.07~\%$ in the modelled light RM and CLV curves during the transit (T2-T3) for the $0.4~{\rm \AA}$ and $0.75~{\rm \AA}$ passbands, respectively. In the observed transmission light curves we measure absorption depths of ${\sim}0.15~\%$ and ${\sim}-0.02~\%$, respectively. 

Near the central time of transit, the observed transmission light curves show a drop in the relative flux. This might be caused by the fact that the S/N of the stellar \ion{Na}{i} lines is very low in the central cores. At zero orbital phase we measure the flux blocked by the planet in the line core, while for shorter and longer orbital phases, this measurement is performed in the wings of the stellar lines, with a higher S/N (because of the orbital motion of the planet). However, this is not reflected in the error bars because they are calculated using the error propagation from the photon noise level of the observed spectra. In the 2D maps from Figure~\ref{fig:TS_comb_zoom} the noise in the central regions of the lines (around $0$~km\,s$^{-1}$) is clearly observed. These maps are shown in the stellar rest frame, but if they are shifted to the planet rest frame (i.e. in the frame we work in), the noisier region moves to different velocities depending on the orbital phases. At zero orbital phase, however, this noise remains at $0$~km\,s$^{-1}$ (i.e. where the measurement is performed).

\subsection{Other lines}
\label{other} Following the same method, we explored other strong lines in the spectrum of HD~209458. Specifically, we analysed the H$\alpha$ line ($6562.81~{\rm \AA}$) using CARMENES and HARPS-N datasets, the \ion{Mg}{i} triplet region ($5167.32$, $5172.68$ and $5183.60~{\rm \AA}$) using HARPS-N, and \ion{K}{i} D1 ($7698.96~{\rm \AA}$) and \ion{Ca}{ii} IRT ($8498.02$, $8542.09$ and $8662.14~{\rm \AA}$) using CARMENES datasets. 

\begin{figure*}[]
\centering
\includegraphics[width=0.95\textwidth]{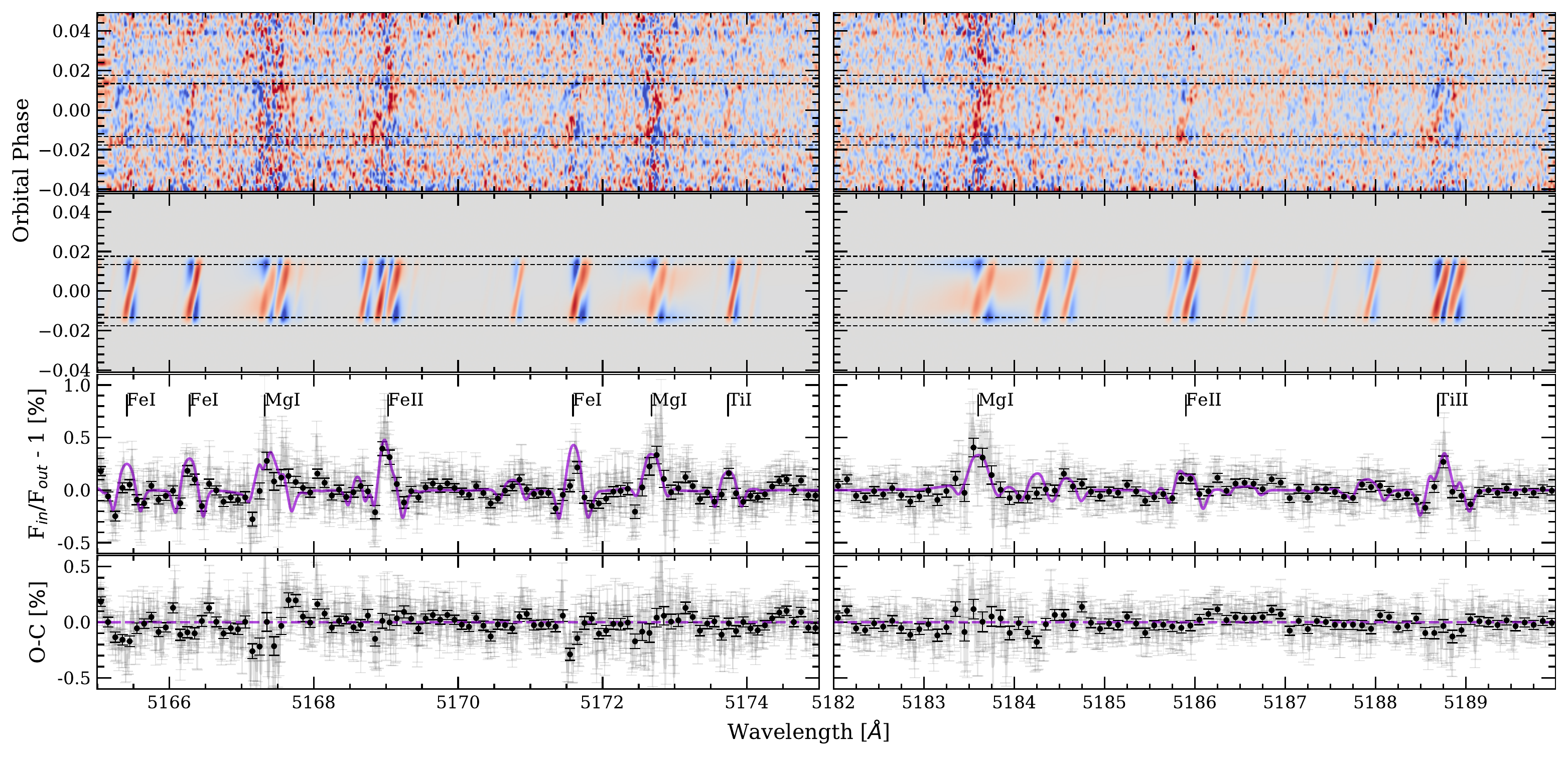}
\caption{Same as Figure~\ref{fig:TS_comb_zoom}, but in the \ion{Mg}{i} triplet ($520~{\rm nm}$) region, obtained with HARPS-N data. The wavelength range is divided into two panels (left and right) for better visualisation. The strongest species are indicated inside the panels. We note that in most of the cases, the effects are not only due to the species specified inside the panel, as the lines are blended. In the second row (starting from top) we additionally include the modelled effects in the stellar rest frame, so that the observed and modelled 2D maps can be easily compared.}
\label{fig:ts_Mg}
\end{figure*}

The effects studied here are especially noticeable around the \ion{Mg}{i} region, from around $5165~{\rm \AA}$ to $5190~{\rm \AA}$. In addition to \ion{Mg}{i}, other strong lines are present in this region, such as \ion{Fe}{i}, \ion{Fe}{ii}, \ion{Ti}{i,} and \ion{Ti}{ii}. The stellar line cores in this region have higher S/N than in the \ion{Na}{i} cores and are not affected by contamination from the Earth atmosphere. Consequently, the combined effects (dominated by the RM effect), which are strong for these lines, are clearly observed (see Figure~\ref{fig:ts_Mg}). In this same figure, we show the observed and modelled 2D maps in the \ion{Mg}{i} region. The effects can be observed in the position shown in the models, and are recovered in the transmission spectrum. For the strongest lines (\ion{Fe}{ii} at $5172~{\rm \AA}$, \ion{Mg}{i} at $5173~{\rm \AA}$, and \ion{Mg}{i} at $5183~{\rm \AA}$), we computed the transmission light curves, which are presented in Fig.~\ref{fig:Fe_TLC} in the appendix.

For H$\alpha$, the transmission spectrum does not show any significant feature. The predicted CLV and RM effects of H$\alpha$ are weak. The measured contrast is around $0.1~\%$ at the line centre. Using CARMENES, we also measured the transmission spectra around the \ion{Ca}{ii} IRT and \ion{K}{i} D1 lines. These transmission spectra are presented in Figures~\ref{fig:ts_Ha}, \ref{fig:ts_K}, and \ref{fig:ts_Ca} in the appendix. In all cases the transmission spectrum shows emission-like features that can be explained by the RM effect. For the \ion{Ca}{ii} IRT lines, the estimated effects do not describe the observations as well as for other lines. One of the reasons, and which would similarly affect H$\alpha$, is that these lines are created in the upper chromosphere and might therefore be affected differently.

\subsection{Systematic effects}
\label{sec:tests}

The error estimates of our measurements come from the propagation of the photon noise through the full analysis, and systematic effects are therefore not taken into account. One way of quantifying the systematic effects is the empirical Monte Carlo (EMC) analysis, presented by \citet{2008Redfield}. The EMC is based on the random selection of individual exposures to build the in-transit and the out-of-transit samples. These samples are then used to compute the transmission spectrum, for which the absorption depth is then measured. In order to have statistical significance, this process was applied $20~000$ times with different random samples. With this, we determined the probability that the measured signal is of planetary origin or is caused by a random combination of the data. This method has been applied in several atmospheric studies such as \citet{2017A&A...602A..36W,2015A&A...577A..62W} and \citet{Jensen2012Ha,jensen2011}.

\begin{figure*}[]
\centering
\includegraphics[width=0.85\textwidth]{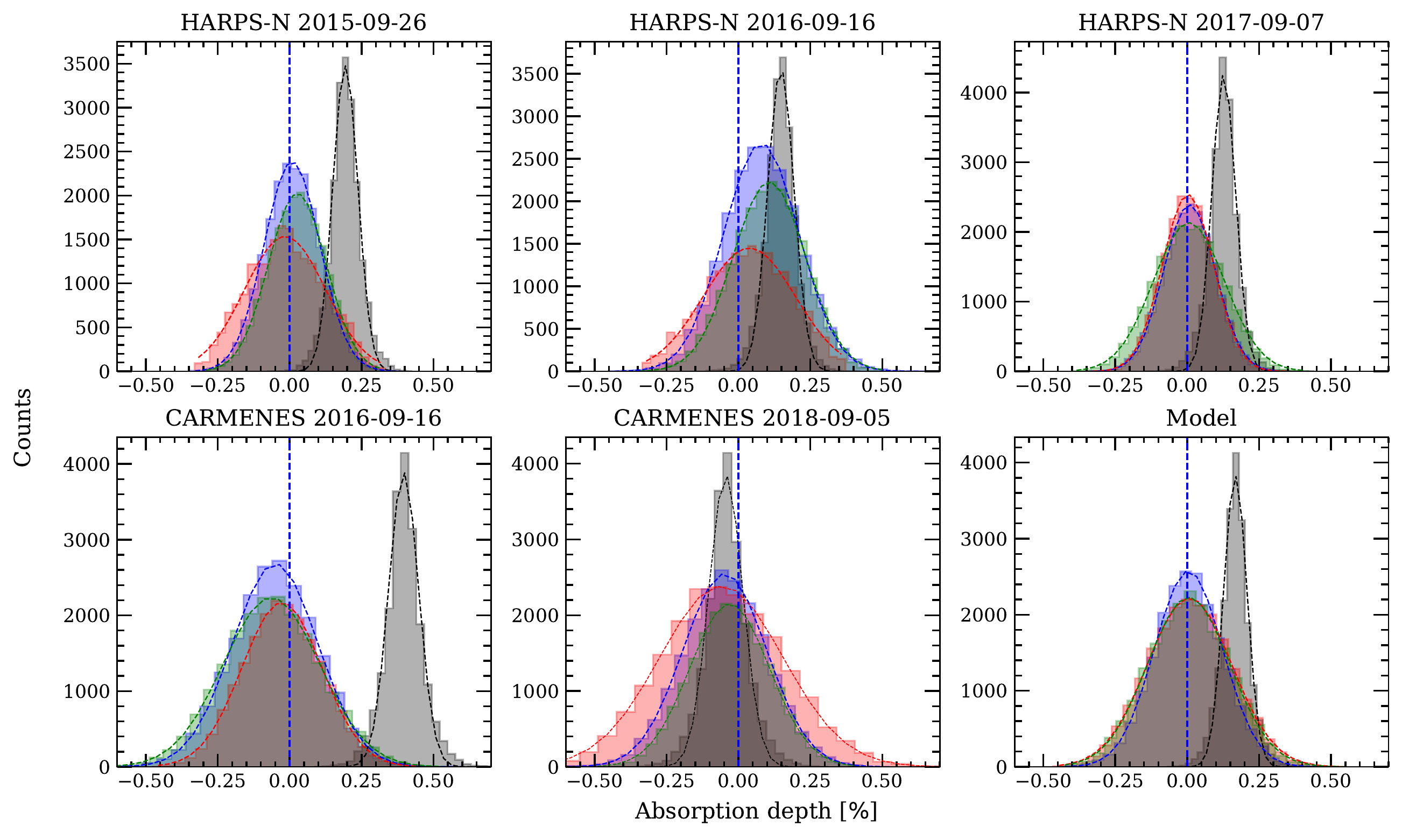}
\caption{Distributions of the EMC analysis of the \ion{Na}{i} lines for the $0.4~{\rm \AA}$ passband. Each individual panel corresponds to the analysis of one night. In green we present the out-out scenario, in red the in-in, in blue the mix-mix, and in grey the in-out. The blue dashed vertical lines show the zero absorption level. We show the Gaussian fit to the histograms as coloured dashed lines.}
\label{fig:EMC}
\end{figure*}

We investigated four different scenarios, three of them described in \citet{2008Redfield}. In summary, the first scenario, called "in-in", takes half of the spectra taken during the transit as the in-transit sample, and the other half as the out-of-transit sample. The second scenario is called "out-out" and takes half of the spectra taken when the planet is not transiting as the in-transit sample and the other half as the out-of-transit sample. At each iteration we randomly selected the spectra that form each sample. The "in-out" scenario is the real case, where the in- and out-of-transit samples correspond to the data taken when the planet is transiting and when it is not, respectively. In this case, the number of spectra in each sample changes in each iteration, but it is always ensured that the number ratio is the same as in the observations and that the smallest number is half the observed in-transit sample. The fourth scenario is called "mix-mix". In this case, the in- and out-of-transit samples contain randomly mixed exposures taken when the planet is and is not transiting. For each scenario, we measured the absorption depth in each of the $20~000$ results. This absorption was measured as the averaged flux inside a passband of $0.4~{\rm \AA}$ centred on the \ion{Na}{i} D2 and D1 lines, respectively. For comparison, the EMC was also applied to the modelled spectra containing both the CLV and RM effects. We added random noise to the model considering a standard deviation of $0.010$, measured in the continuum of HARPS-N normalised data of night 5, and we used the same number of in- and out-of-transit exposures as in this particular night. The histograms with the absorption depth measured for the $20~000$ cases are shown in Fig.~\ref{fig:EMC}.

In all cases, the three control scenarios (in-in, out-out, and mix-mix) present distributions that are centred around $0~\%$ of absorption. The in-out scenario presents distributions centred at $0.19\pm0.04~\%$, $0.15\pm0.05~\%$, and $0.13\pm0.04~\%$ for the three HARPS-N nights. For the two nights observed with CARMENES, these distributions are centred at $0.39\pm0.05~\%$ and $-0.04\pm0.06~\%$. On the other hand, the model shows an in-out distribution centred at $0.17\pm0.04~\%$. The error bars of the values correspond to the standard deviation of the distributions (see the absorption depth values summarised in Table~\ref{Tab:EMC_values}).

\begin{table}[]
\centering
\caption{Absorption depth (in $\%\text{}$) measured on the EMC distributions computed using the $0.4~{\rm \AA}$ passband.}
\resizebox{0.45\textwidth}{!}{\begin{tabular}{lcccc}
\hline\hline
 & in-out & in-in & out-out & mix-mix\\ \hline
\\[-1em]
Night 1 & $0.19\pm0.04$ & $-0.01\pm0.14$ & $0.03\pm0.10$ & $0.01\pm0.10$\\ 
\\[-1em]
Night 3 & $0.15\pm0.05$ & $0.04\pm0.16$ & $0.11\pm0.13$ & $0.08\pm0.13$ \\ 
\\[-1em]
Night 5 & $0.13\pm0.04$ & $0.00\pm0.09$ & $0.01\pm0.12$ & $0.01\pm0.09$\\ 
\\[-1em]
Night 7 & $0.39\pm0.05$ &  $-0.03\pm0.14$ & $-0.06\pm0.17$ & $-0.05\pm0.15$\\ 
\\[-1em]
Night 9 & $-0.04\pm0.06$ & $-0.06\pm0.21$& $-0.03\pm0.14$ & $-0.05\pm0.14$\\ 
\\[-1em]
Model  & $0.17\pm0.04$ &  $0.01\pm0.15$ & $0.00\pm0.14$ & $0.00\pm0.12$\\
\\[-1em]
\hline
\end{tabular}}\\
\label{Tab:EMC_values}
\end{table}

Although we observe that the S/N in the line cores is very low as a result of the deep stellar lines, in contrast with the control distributions (which are centred at $0\%$), the in-out distributions are centred at a positive absorption depth, as measured in the transmission spectra. This does not occur in the case of the CARMENES night 9 observation, for which the in-out distribution is centred near $\sim0\%$. In the individual transmission spectrum (see right panel of Figure~\ref{fig:indiv_TS_CARM}) a drop in flux can be observed at the left side of the laboratory position for both \ion{Na}{i} D2 and D1 lines because the spectra are noisier. In the $0.4~{\rm \AA}$ passband, this region is partially included and decreases the absorption depth.

\section{Comparison with previous results}
\label{sec:comp}

HD~209458b is one of the most frequently studied planets, with several detections of the \ion{Na}{i} doublet using different facilities. Here, we compare our results around these spectral lines with those presented in \citet{Sing2008HD209ApJ...686..658S}, \citet{2008SnellenHD209}, and \citet{Simon2009IAUS..253..520A}.

\begin{figure}[]
\centering
\includegraphics[width=0.48\textwidth]{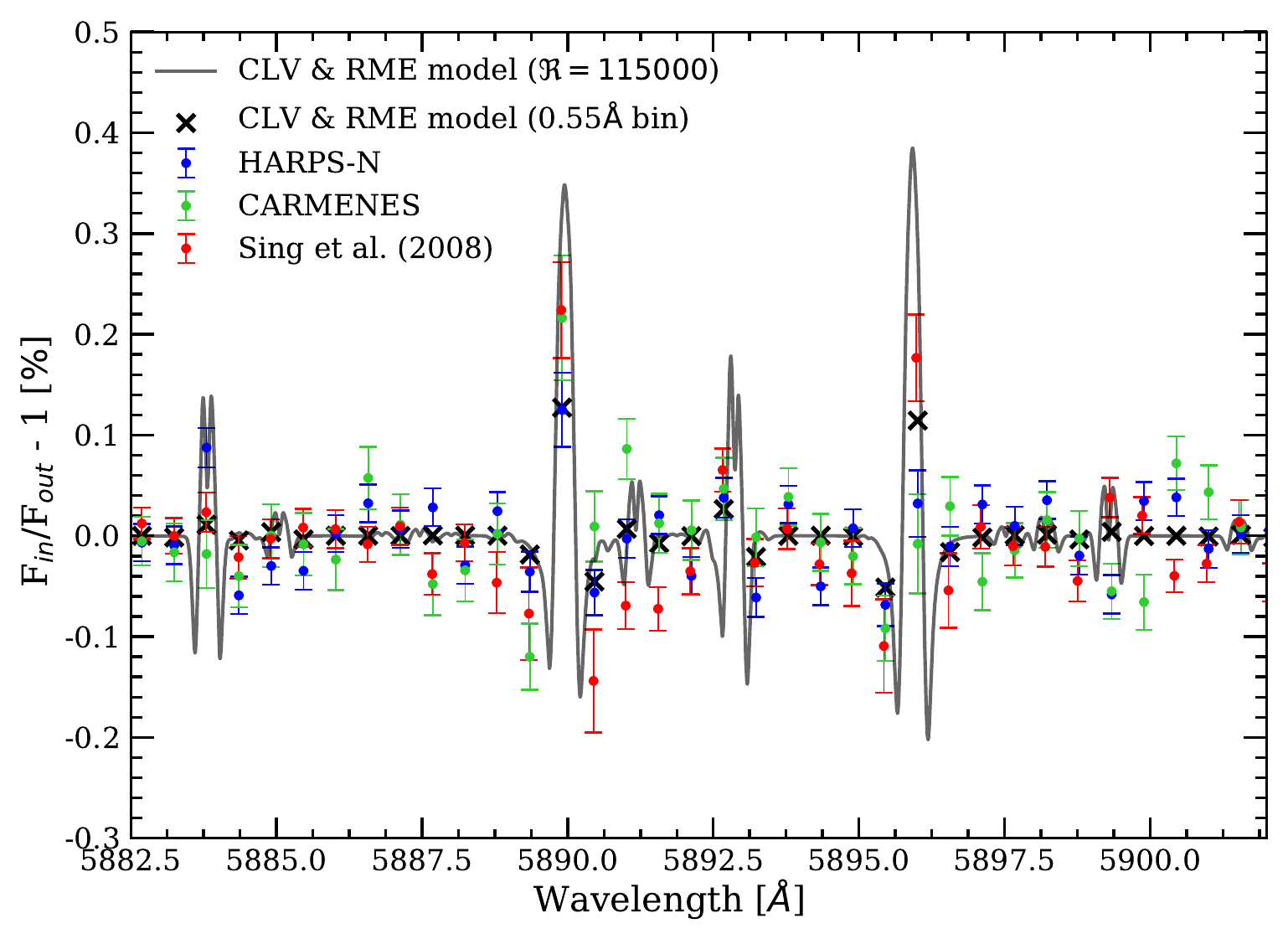}
\includegraphics[width=0.49\textwidth]{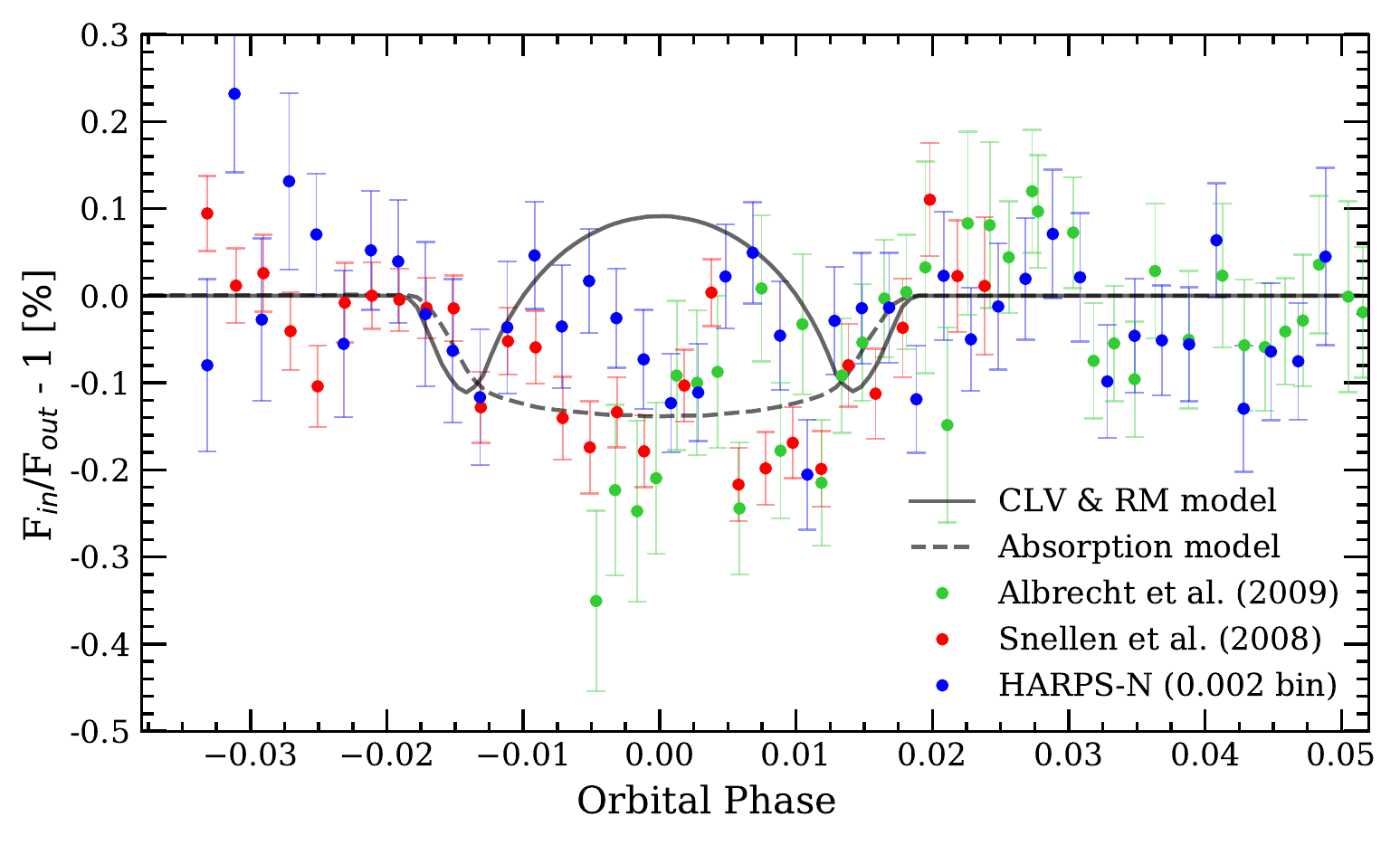}
\caption{Top panel: Comparison of the CARMENES and HARPS-N transmission spectra with the result presented in \citet{Sing2008HD209ApJ...686..658S}. The CARMENES and HARPS-N results are presented using a binning width of $0.55~{\rm \AA}$. Bottom panel: HD~209458b transmission light curves from \citet{2008SnellenHD209}, \citet{Simon2009IAUS..253..520A}, and this work (binned by $0.002$ in orbital phase). The dashed line shows the transmission model as presented in those studies, while the solid line is the expected CLV and RM effects obtained with the same method.}
\label{fig:Comparing}
\end{figure}

We compared our results with the transmission spectrum obtained by \citet{Sing2008HD209ApJ...686..658S} using mid-resolution ($\Re = 5540$) observations with the STIS at HST (see Figure~\ref{fig:Comparing}). For this comparison, we binned the transmission spectra from Figure~\ref{fig:TS_comb} in $0.55~{\rm \AA}$ intervals, which is the  STIS pixel size. The centre of each bin was located at the positions presented in \citet{Sing2008HD209ApJ...686..658S}. The modelled spectrum was computed using only the orbital phases covered by HST, and was then binned at the same intervals as the data. Our high-resolution binned data and our model are consistent with the HST observations. \citet{Sing2008HD209ApJ...686..658S} discarded the two points falling on the \ion{Na}{i} line cores from their analysis because of possible telluric contamination: their wavelength positions were consistent with the Earth's radial velocities during the observations. However, our results reveal that an alternative explanation is the combination of the RM+CLV effects, which is also valid for other wavelengths such as the \ion{Ni}{i} position at $5893~{\rm \AA}$, which is also visible in the figure. In HST observations, this telluric contamination is absent in other strong lines such as \ion{Fe}{i}, while the CLV and RM effects should remain (see Figure~\ref{fig:ts_Mg}), although they are perhaps not detectable by STIS. When we compared the results, we noted an offset of $-0.2~{\rm \AA}$ (half of the STIS pixel size) between HST and the results presented here.

 \citet{2008SnellenHD209} and \citet{Simon2009IAUS..253..520A} measured the transmission light curves by integrating the flux inside the stellar line cores, in the stellar rest frame, and compared the flux with the one measured inside two reference passbands. They measured a \ion{Na}{i} absorption of $0.135\pm0.017~\%$ with a passband of $0.75~{\rm \AA}$. Following this same method, we built a transmission light curve with the HARPS-N data. In Figure~\ref{fig:Comparing} we compare our own light curve with those presented in \citet{2008SnellenHD209} and \citet{Simon2009IAUS..253..520A}. None of the results agrees with the modelled CLV and RM effects light curve. On the other hand, although we are not able to reproduce the full transmission light curve from their results with our data, most of the points are consistent considering the error bars. Light curve observations with higher S/N\ might help solve these issues. Nevertheless, the agreement between our models and the transmission spectrum results obtained in Sect.~\ref{sec:results} (Figures~\ref{fig:TS_comb}, \ref{fig:TS_comb_zoom}, and \ref{fig:ts_Mg}) gives us confidence that the signals seen in our analysis can be explained without invoking planetary absorption.

Finally, \citet{Keles2019MNRAS.489L..37K} studied the presence of \ion{K}{i} in the atmosphere of HD~209458b. Interestingly, they reported an emission-like behaviour at low bandwidths, which would be consistent with the RM effect and non-detection of \ion{K}{i} line at $7699~{\rm \AA}$ reported in this work.

\section{Discussion and conclusions}
\label{sec:disc}

We combined nine transit observations (five with HARPS-N and four with CARMENES) to measure the RM effect using the radial velocity measurements obtained from SERVAL. The best-fit model shows a projected spin-orbit angle of $\lambda=-1.6\pm0.3$, indicating a well-aligned orbit, as presented by \citet{Winn2005HD209} and \citet{Albrecht2012ApJ...757...18A}. Because of its brightness and the RM amplitude, HD~209458b is an excellent planet for chromatic RM studies (\citealt{SnellenChromaticRM}, \citealt{DiGloria2015A&A...580A..84D}, and \citealt{Yan2015ApJ...806L..23Y}), especially using ESPRESSO-like observations with high S/N.

We also computed the transmission spectrum of HD~209458b around the \ion{Na}{i} using three archival nights observed with HARPS-N, and two with CARMENES. The importance of considering the CLV and RM effect in atmospheric studies has previously been pointed out. \citet{Yan2017A&A...603A..73Y} predicted the significance of the CLV effect in atmospheric studies of HD~209458b around \ion{Na}{i}. In particular, we observe that this effect becomes significant in the transmission light curves ($\sim0.1\%$ of contrast, see Figure~\ref{fig:mod_TS_TLC}). Here, in addition to this effect, we also considered the RM effect contribution, which is found to strongly increase the residual effects in the line cores of the transmission spectrum by a factor of about four when computed in the planet rest frame. For all individual nights, the resulted transmission spectra of the exoplanetary atmosphere show an emission-like signal instead of an expected absorption signal, as has been found in previous studies by \citet{Simon2009IAUS..253..520A}, \citet{2008SnellenHD209}, \citet{Sing2008HD209ApJ...686..658S}, and \citet{2002ApJ...568..377C}. The transmission spectra presented here are consistent with the modelled CLV and RM effects on the stellar line profiles without considering any contribution from the exoplanet atmosphere.

The same is observed in the transmission light curves that are computed using narrow passbands, where the positive contribution of the RM effect is the dominating factor. For wide passbands, however, the effects are diluted and the transmission light curves do not present enough S/N to observe any clear behaviour. Further observations are needed at higher S/N to build reliable light curves that can be compared to our models.

Finally, we compared our measurements and models to previous studies of the presence of \ion{Na}{i} in the atmosphere of HD~209458b. Our results reveal that an alternative explanation for the transmission spectrum derived from \textit{HST} observations is the combination of the RM and CLV effects. When we compared this to High-Dispersion Spectrograph/Subaru observations, we were unable to reproduce the full transmission light curve from their results with our data, but most of the points are consistent considering the error bars. Moreover, none of the light curves are consistent with the modelled effects. Light curve observations with higher S/N are needed to solve this issue. Nevertheless, the agreement between our models and the transmission spectrum results obtained in Sect.~\ref{sec:results} (Figures~\ref{fig:TS_comb}, \ref{fig:TS_comb_zoom} and \ref{fig:ts_Mg}) gives us confidence that the signals seen in all datasets can be explained without invoking planetary absorption by \ion{Na}{i} in the atmosphere of HD~209458b. Our results also imply that the detection of atmospheric features needs to account for these effects. Detailed modelling of both RM and CLV effects like this is mandatory when the characterisation of small Earth-like planets around low-mass stars is attempted with the ELTs in the coming decades.


\begin{acknowledgements}
 Based on observations made with the Italian Telescopio Nazionale Galileo (TNG) operated on the island of La Palma by the Fundación Galileo Galilei of the INAF (Istituto Nazionale di Astrofisica) at the Spanish Observatorio del Roque de los Muchachos of the Instituto de Astrofisica de Canarias.
 
CARMENES is an instrument for the Centro Astronómico Hispano-Alemán de Calar Alto (CAHA, Almería, Spain). CARMENES is funded by the German Max-Planck-Gesellschaft (MPG), the Spanish Consejo Superior de Investigaciones Científicas (CSIC), the European Union through FEDER/ERF FICTS-2011-02 funds, and the members of the CARMENES Consortium (Max- Planck-Institut für Astronomie, Instituto de Astrofísica de Andalucía, Landessternwarte Königstuhl, Institut de Ciències de l'Espai, Insitut für Astrophysik Göttingen, Universidad Complutense de Madrid, Thüringer Landessternwarte Tautenburg, Instituto de Astrofísica de Canarias, Hamburger Sternwarte, Centro de Astrobiología and Centro Astronómico Hispano- Alemán), with additional contributions by the Spanish Ministry of Economy, the German Science Foundation through the Major Research Instrumentation Programme and DFG Research Unit FOR2544 “Blue Planets around Red Stars”, the Klaus Tschira Stiftung, the states of Baden-Württemberg and Niedersachsen, and by the Junta de Andalucía.

 This work is partly financed by the Spanish Ministry of Economics and Competitiveness through project ESP2016-80435-C2-2-R and ESP2017-87143-R. G. C. acknowledges the support by the Natural Science Foundation of Jiangsu Province (Grant No. BK20190110), the National Natural Science Foundation of China (Grant No. 11503088, 11573073, 11573075). F.Y. acknowledges the support of the DFG priority program SPP 1992 "Exploring the Diversity of Extrasolar Planets (RE 1664/16-1). This work made use of PyAstronomy and of the VALD database, operated at Uppsala University, the Institute of Astronomy RAS in Moscow, and the University of Vienna.

\end{acknowledgements}

%
%
\bibliographystyle{aa.bst} 
\bibliography{aa.bib} 





   
  



%

%
\onecolumn
\begin{appendix} 

\section{MCMC results and probability distributions}

\begin{figure*}[h]
\centering
\includegraphics[width=0.98\textwidth]{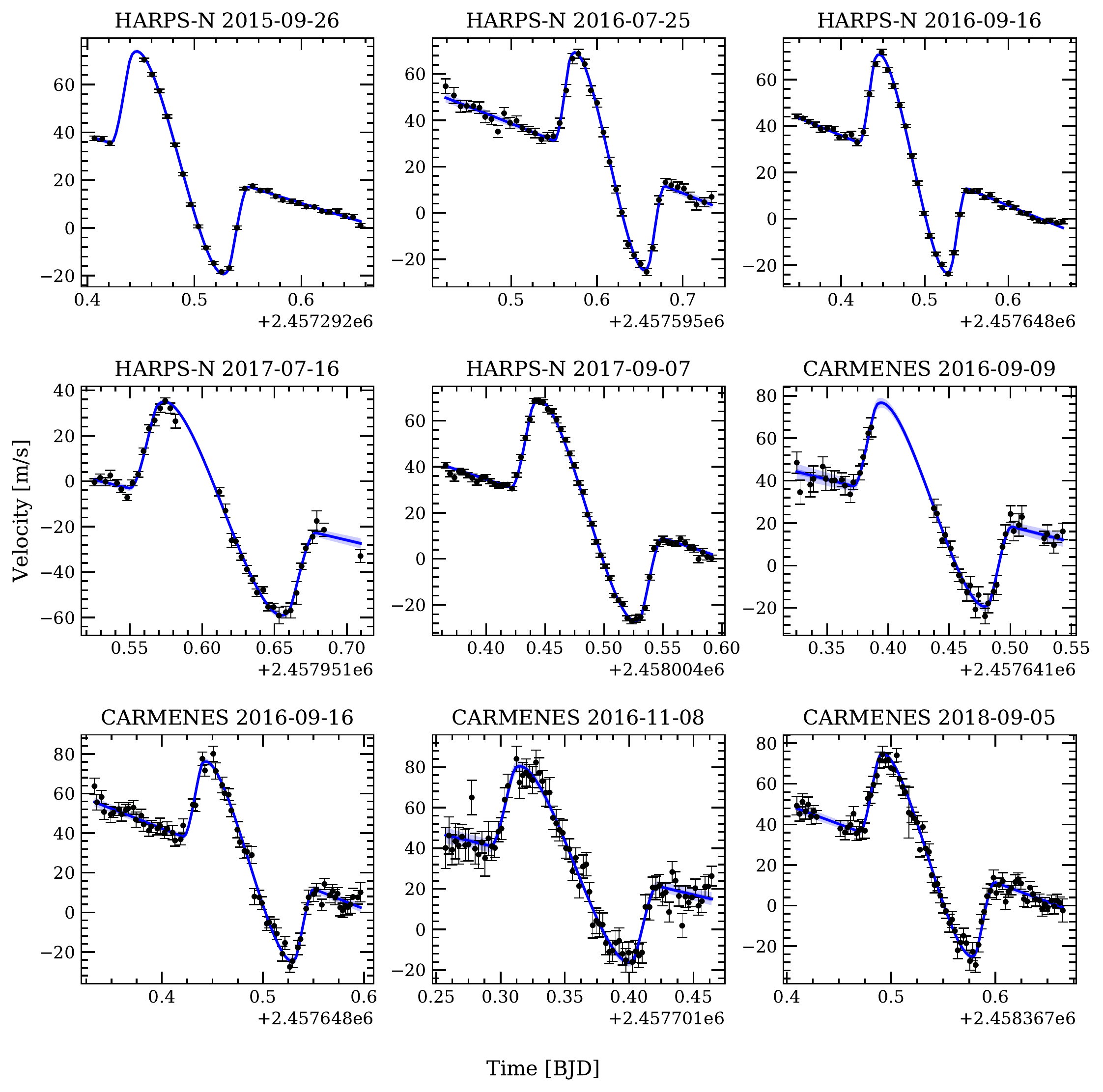}
\caption{Measured stellar radial velocities of HD~209458 during the transit (black) for different nights and instruments. In blue we show the best-fit model obtained with the MCMC procedure.}
\label{fig:RM_indiv}
\end{figure*}

\newpage 

\begin{table}[]
\centering
\caption{Extension of Table~\ref{tab:RM_res}. MCMC best-fit $\Delta v$ and $K_{\star}$ values with $1\sigma$, and $3\sigma$ error bars for different assumptions: $i_{\star}=90~\deg$ and circular orbit (Case~1), $i_{\star}$ free and circular orbit (Case~2), and $i_{\star}=90~\deg$ and free eccentricity (Case~3).}
\begin{tabular}{ll|lll|lll|lll}
\hline\hline
& & & Case 1 & & & Case 2 & & & Case 3 & \\ \hline
\\[-1em]
Parameter & Units & Value & $1\sigma$ & $3\sigma$ & Value & $1\sigma$ & $3\sigma$ & Value & $1\sigma$ & $3\sigma$\\ \hline
\\[-1em]
$\Delta v^{\rm H1}$ & m\,s$^{-1}$ & $26.4$ & $\pm0.2$ & $\pm0.6$&$26.4$&$\pm 0.2$ &$^{+0.6}_{-0.7}$&$25.4$&$^{+6.0}_{-7.5}$&$^{+17.8}_{-18.8}$\\  
\\[-1em]
$\Delta v^{\rm H2}$ & m\,s$^{-1}$ & $21.4$ & $\pm0.3$ & $\pm0.9$&$21.4$&$\pm0.3$&$^{+1.0}_{-0.8}$&$20.4$&$^{+6.4}_{-8.0}$&$^{+18.3}_{-19.9}$\\ 
\\[-1em]
$\Delta v^{\rm H3}$ & m\,s$^{-1}$ & $22.9$ & $\pm0.2$ & $^{+0.5}_{-0.4}$&$22.9$&$\pm0.2$&$\pm0.4$&$21.9$&$^{+6.4}_{-8.0}$&$^{+19.0}_{-20.1}$\\ 
\\[-1em]
$\Delta v^{\rm H4}$ & m\,s$^{-1}$ & $-12.9$ & $\pm0.3$ & $\pm1.0$&$-12.9$&$\pm0.4$&$\pm1.0$&$-13.8$&$^{+6.3}_{-8.1}$&$^{+18.9}_{-20.0}$\\ 
\\[-1em]
$\Delta v^{\rm H5}$ & m\,s$^{-1}$ & $20.0$ & $\pm0.2$ & $\pm0.5$&$20.0$&$\pm0.2$&$\pm0.5$&$18.9$&$^{+7.2}_{-9.1}$&$^{+20.7}_{-22.5}$\\ 
\\[-1em]
$\Delta v^{\rm C1}$ & m\,s$^{-1}$ & $27.9$ & $\pm0.6$ & $^{+1.7}_{-1.8}$&$27.9$&$\pm0.6$&$^{+1.8}_{-1.7}$&$26.8$&$^{+6.5}_{-8.1}$&$^{+18.1}_{-20.2}$\\ 
\\[-1em]
$\Delta v^{\rm C2}$ & m\,s$^{-1}$ & $24.9$ & $\pm0.4$ & $\pm1.2$&$24.9$&$\pm0.4$&$^{+1.3}_{-1.1}$&$23.5$&$^{+8.7}_{-10.9}$&$^{+24.9}_{-27.7}$\\ 
\\[-1em]
$\Delta v^{\rm C3}$ & m\,s$^{-1}$ & $31.2$ & $\pm0.7$ & $^{+2.0}_{-1.9}$&$31.1$&$\pm0.6$&$^{+1.9}_{-1.7}$&$30.4$&$^{+6.6}_{-8.2}$&$^{+19.2}_{-21.7}$\\ 
\\[-1em]
$\Delta v^{\rm C4}$ & m\,s$^{-1}$ & $24.0$ & $\pm0.4$ & $\pm1.2$&$24.0$&$\pm0.4$&$\pm1.1$&$22.7$&$^{+8.1}_{-10.1}$&$^{+24.2}_{-25.9}$\\ 
\\[-1em]
\hline 
\\[-1em]
$K_{\star}^{\rm H1}$ & m\,s$^{-1}$ & $79.6$ & $\pm1.3$ & $^{+3.8}_{-3.6}$&$79.8$&$^{+1.2}_{-1.3}$&$^{+3.7}_{-3.8}$ & $71.9$&$^{+5.5}_{-6.5}$&$^{+9.9}_{-11.6}$\\ 
\\[-1em]

$K_{\star}^{\rm H2}$ & m\,s$^{-1}$ & $84.9$ & $\pm2.1$ & $^{+6.0}_{-6.1}$ & $84.0$&$^{+2.1}_{-2.2}$&$^{+5.3}_{-6.7}$ & $77.1$&$^{+5.7}_{-7.1}$&$^{+11.1}_{-14.8}$\\ 
\\[-1em]
$K_{\star}^{\rm H3}$ & m\,s$^{-1}$ &$85.2$ & $\pm0.8$ & $\pm2.4$ & $85.4$&$^{+0.8}_{-0.9}$&$^{+2.4}_{-2.6}$ & $77.2$&$^{+5.8}_{-7.0}$&$^{+10.3}_{-13.0}$\\ 
\\[-1em]
\\[-1em]
$K_{\star}^{\rm H4}$ & m\,s$^{-1}$ &$85.5$ & $^{+3.5}_{-3.4}$ & $^{+10.2}_{-9.8}$ & $84.5$&$^{+3.7}_{-3.4}$&$^{+10.0}_{-9.5}$ & $77.2$&$^{+6.4}_{-7.6}$&$^{+14.0}_{-15.5}$\\ 
\\[-1em]
\\[-1em]
$K_{\star}^{\rm H5}$ & m\,s$^{-1}$ &$95.9$ & $\pm1.4$ & $^{+3.7}_{-4.1}$ & $95.5$&$^{+1.4}_{-1.3}$&$^{+4.0}_{-3.9}$ & $86.7$&$^{+7.0}_{-7.9}$&$^{+11.8}_{-15.5}$\\
\\[-1em]
\\[-1em]
$K_{\star}^{\rm C1}$ & m\,s$^{-1}$ &$83.1$ & $^{+5.5}_{-5.2}$ & $^{+14.9}_{-15.0}$ & $85.9$&$\pm5.2$&$^{+15.0}_{-15.5}$ & $75.3$&$^{+7.8}_{-7.9}$&$^{+20.4}_{-15.1}$\\ 
\\[-1em]
\\[-1em]
$K_{\star}^{\rm C2}$ & m\,s$^{-1}$ &$114.5$ & $\pm2.9$ & $^{+8.5}_{-8.0}$ & $113.9$&$\pm2.9$&$^{+6.0}_{-8.6}$ & $103.4$&$^{+8.4}_{-9.5}$&$^{+15.8}_{-20.3}$\\ 
\\[-1em]
\\[-1em]
$K_{\star}^{\rm C3}$ & m\,s$^{-1}$ &$85.8$ & $^{+6.0}_{-6.2}$ & $^{+13.7}_{-18.2}$ & $79.5$&$\pm4.9$&$^{+19.7}_{-13.7}$ & $78.1$&$^{+8.6}_{-8.2}$&$^{+22.7}_{-17.0}$\\ 
\\[-1em]
\\[-1em]
$K_{\star}^{\rm C4}$ & m\,s$^{-1}$ &$108.0$ & $^{+3.1}_{-3.0}$ & $^{+8.7}_{-8.9}$ & $109.1$&$^{+3.0}_{-3.1}$&$^{+8.4}_{-8.9}$ & $97.4$&$^{+7.3}_{-8.9}$&$^{+15.8}_{-18.6}$\\
\\[-1em]
\\[-1em]
\hline
\end{tabular}\\
\label{tab:RM_res_add}
\end{table}

\begin{figure}[h]
\centering
\includegraphics[width=0.55\textwidth]{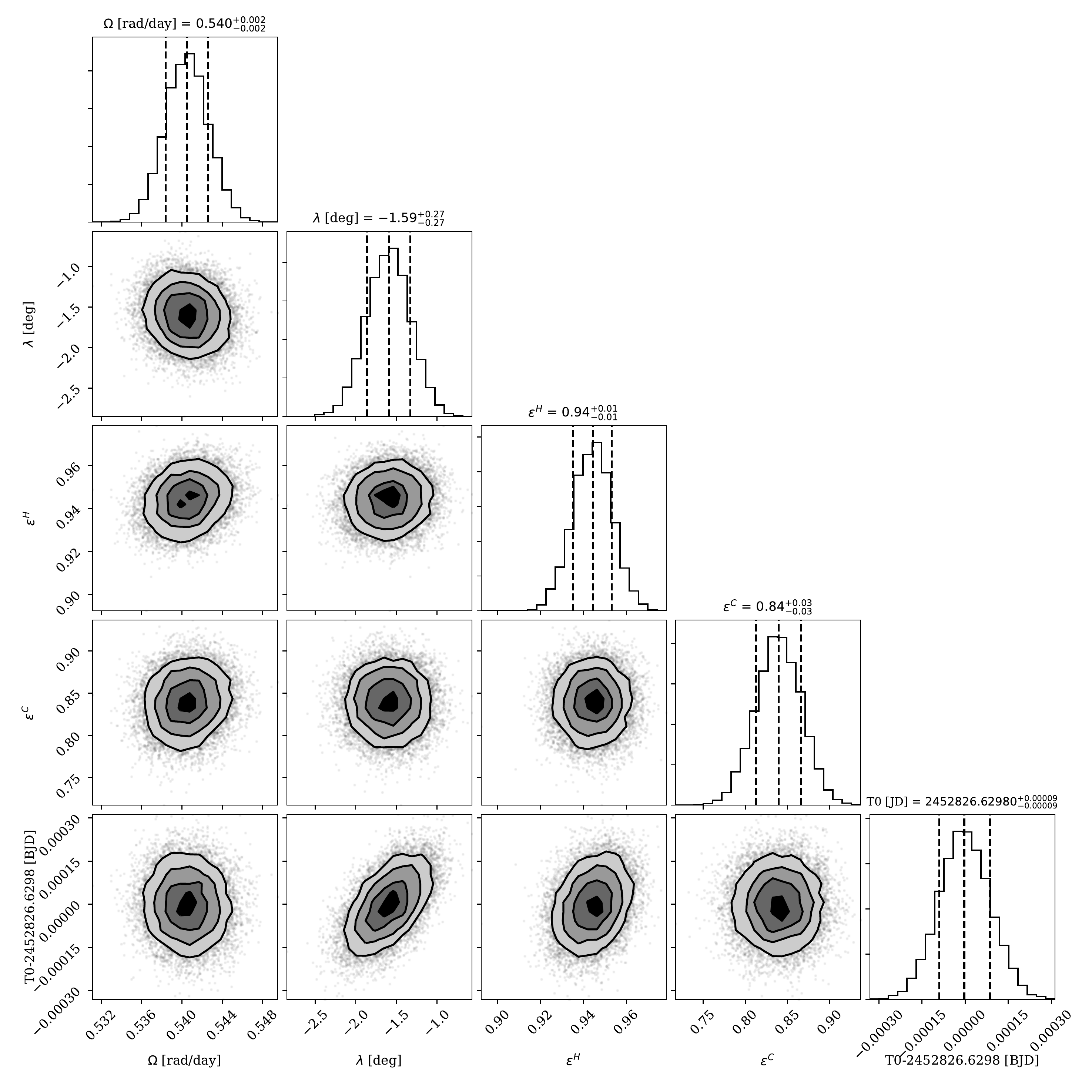}
\caption{Correlation diagrams for the probability distribution of the RM model parameters for HD~209458b, obtained assuming $i_{\star} = 90~\deg$ and circular orbit (Case~1). The dashed lines overimposed on the histograms correspond to the 16 and 84 percentiles used to obtain the $1\sigma$ statistical errors. One hundred walkers and $10^5$ steps are used in this analysis.}
\label{fig:RM_corner}
\end{figure}

\newpage

\section{Individual transmission spectra}
\subsection{HARPS-N data}
\begin{figure}[h]
\centering
\includegraphics[width=0.49\textwidth]{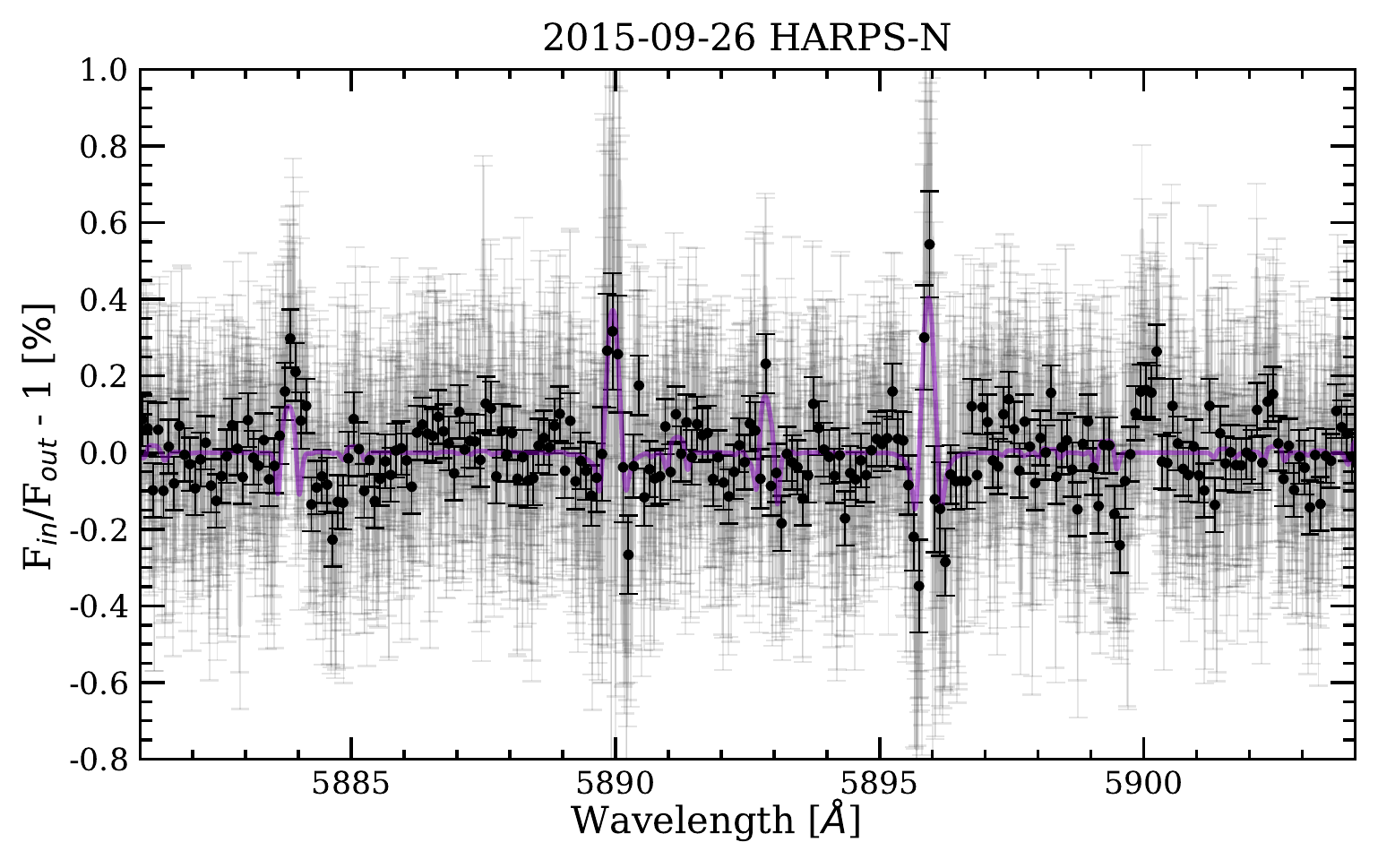}
\includegraphics[width=0.49\textwidth]{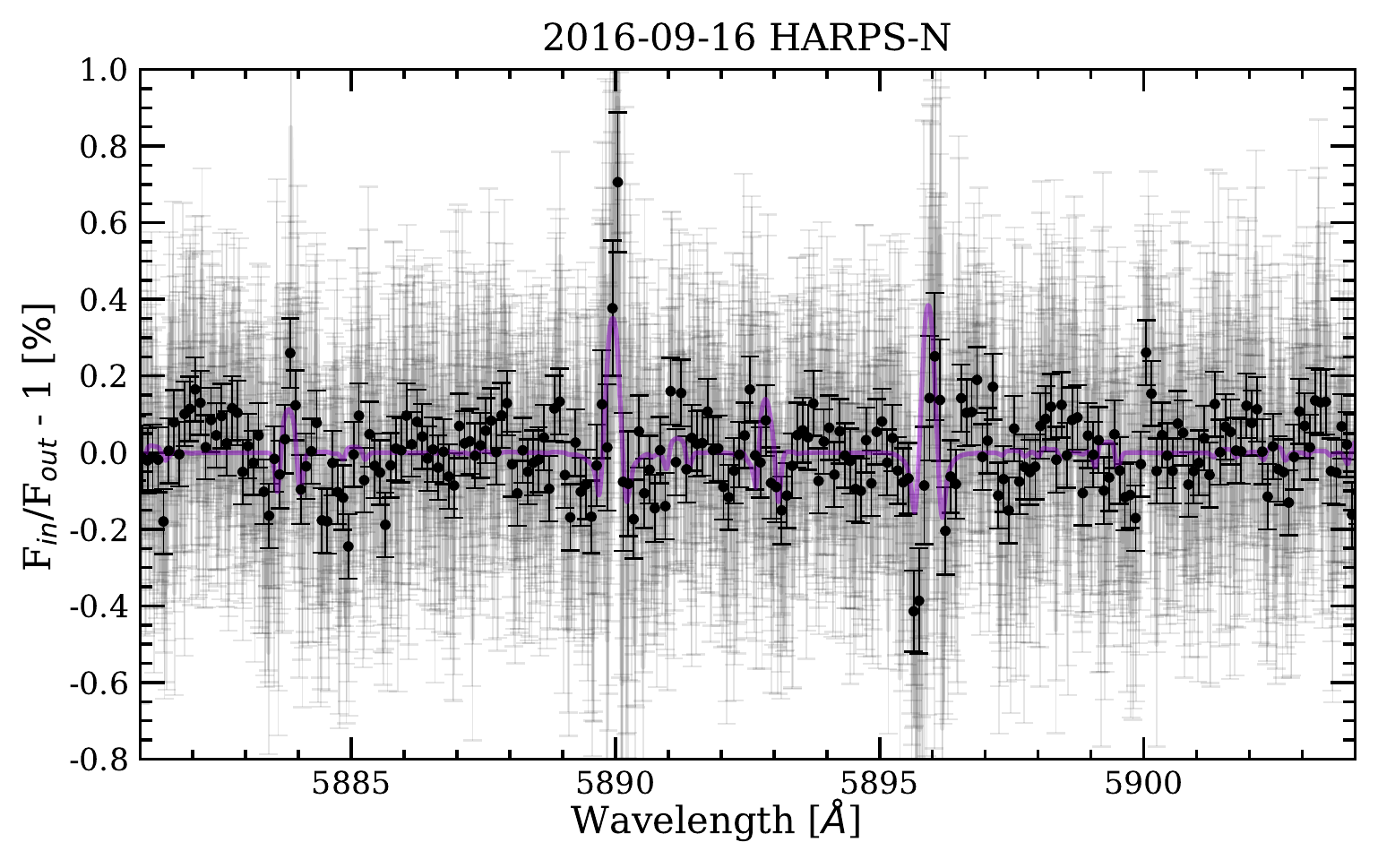}
\includegraphics[width=0.49\textwidth]{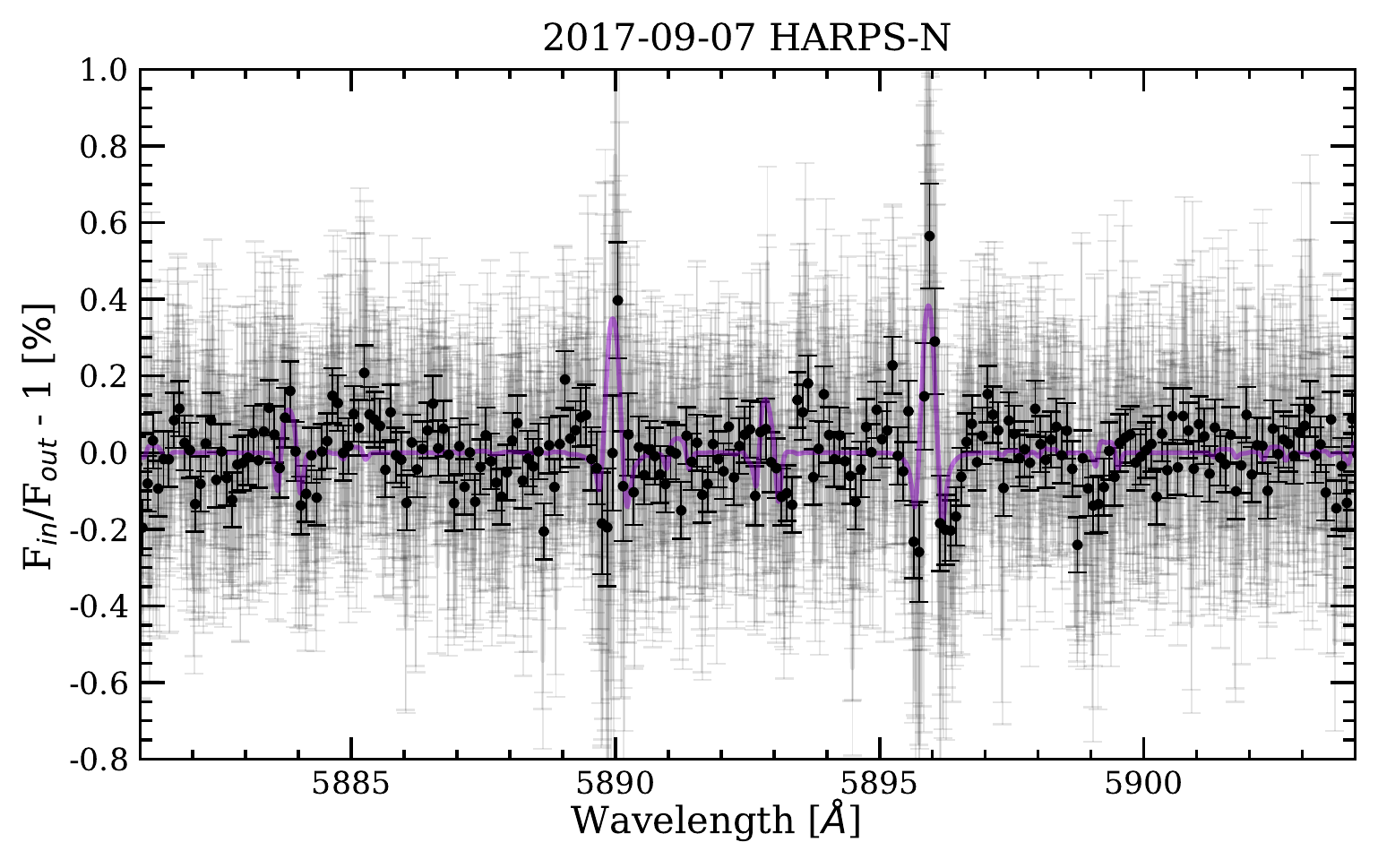}
\caption{HD~209458b transmission spectra around the \ion{Na}{i} doublet for three different HARPS-N data sets. In light grey we show the original data, and as black dots the data binned by $0.1~\mathrm{\AA}$. In purple we show the RME+CLV model for each data set.}
\label{fig:indiv_TS_HARPS}
\end{figure}

\subsection{CARMENES data}
\begin{figure}[h]
\centering
\includegraphics[width=0.49\textwidth]{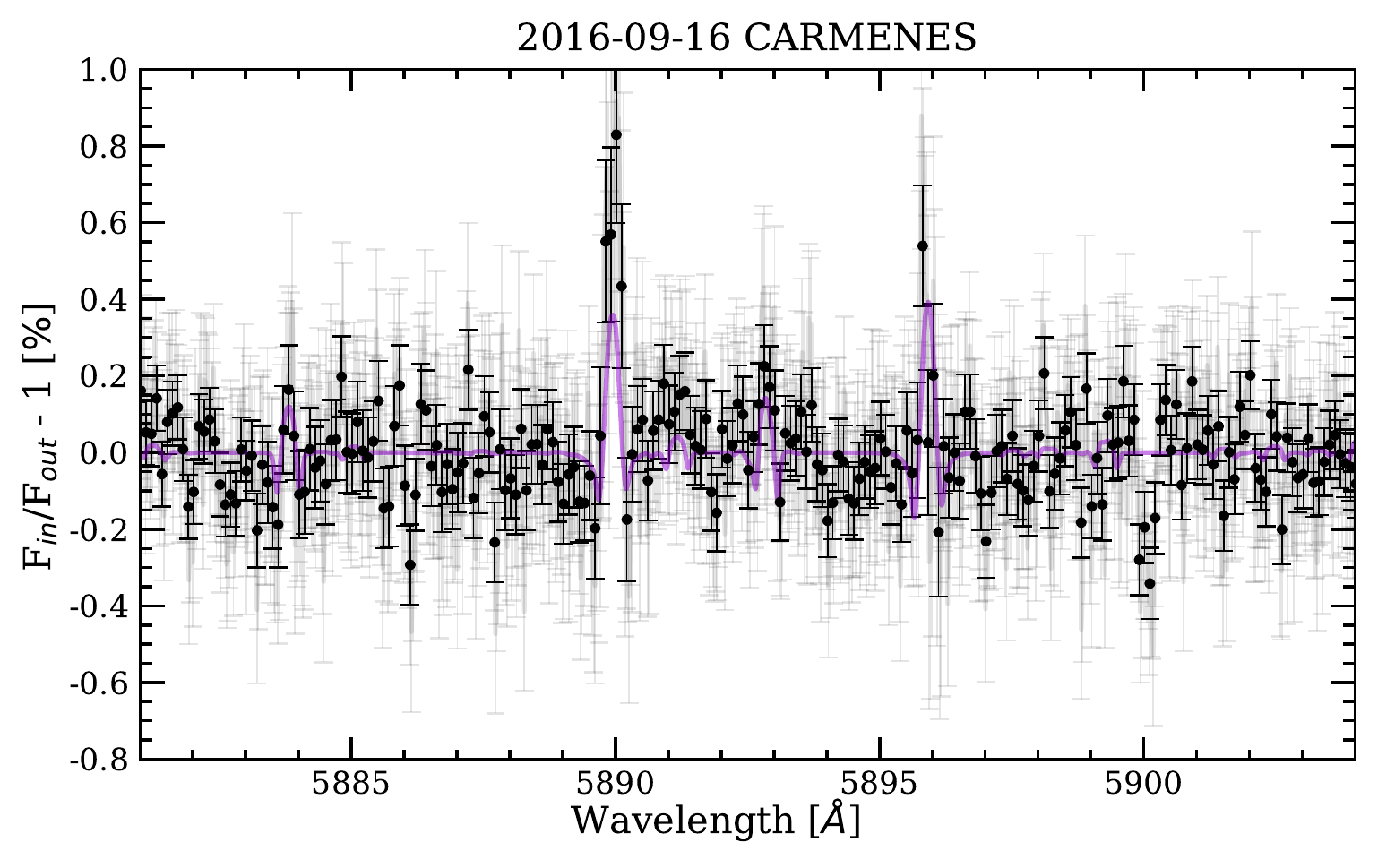}
\includegraphics[width=0.49\textwidth]{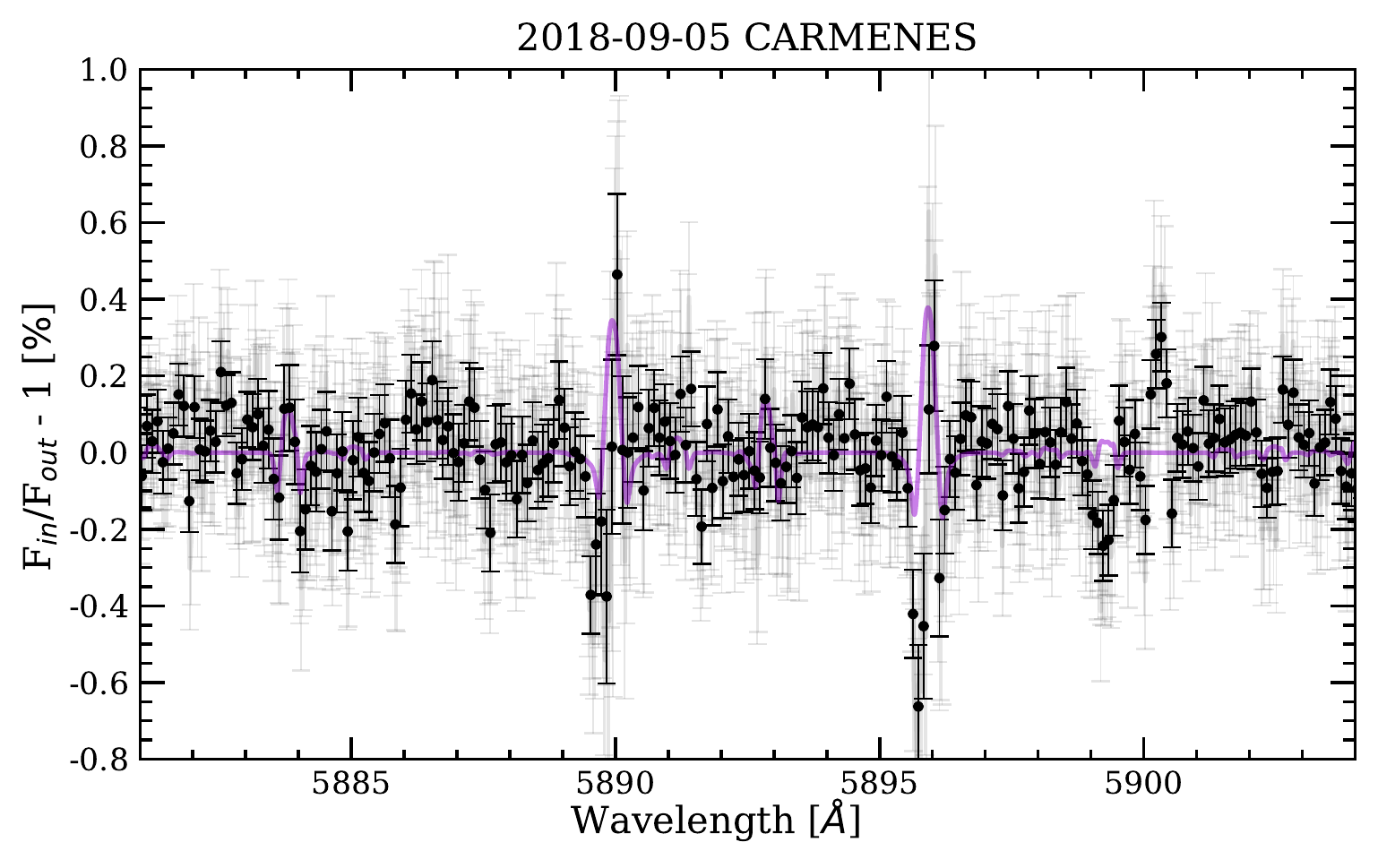}
\caption{Same as Figure~\ref{fig:indiv_TS_HARPS}, but for CARMENES data.}
\label{fig:indiv_TS_CARM}
\end{figure}

\newpage

\section{Individual instrument and \ion{Na}{i} lines transmission light curves}

\label{ap:TLC}
\begin{figure}[h]
\centering
\includegraphics[width=0.3\textwidth]{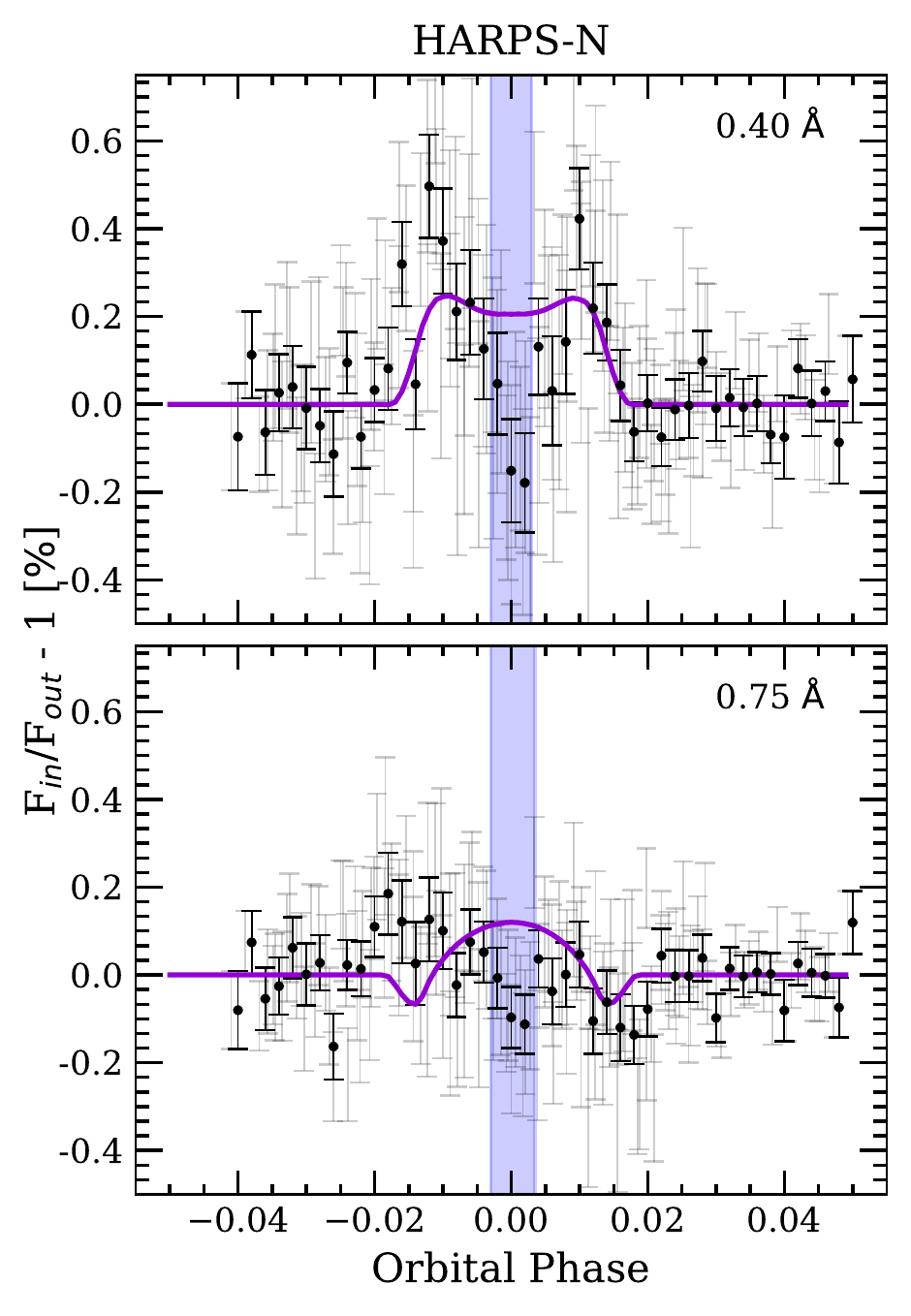}
\includegraphics[width=0.3\textwidth]{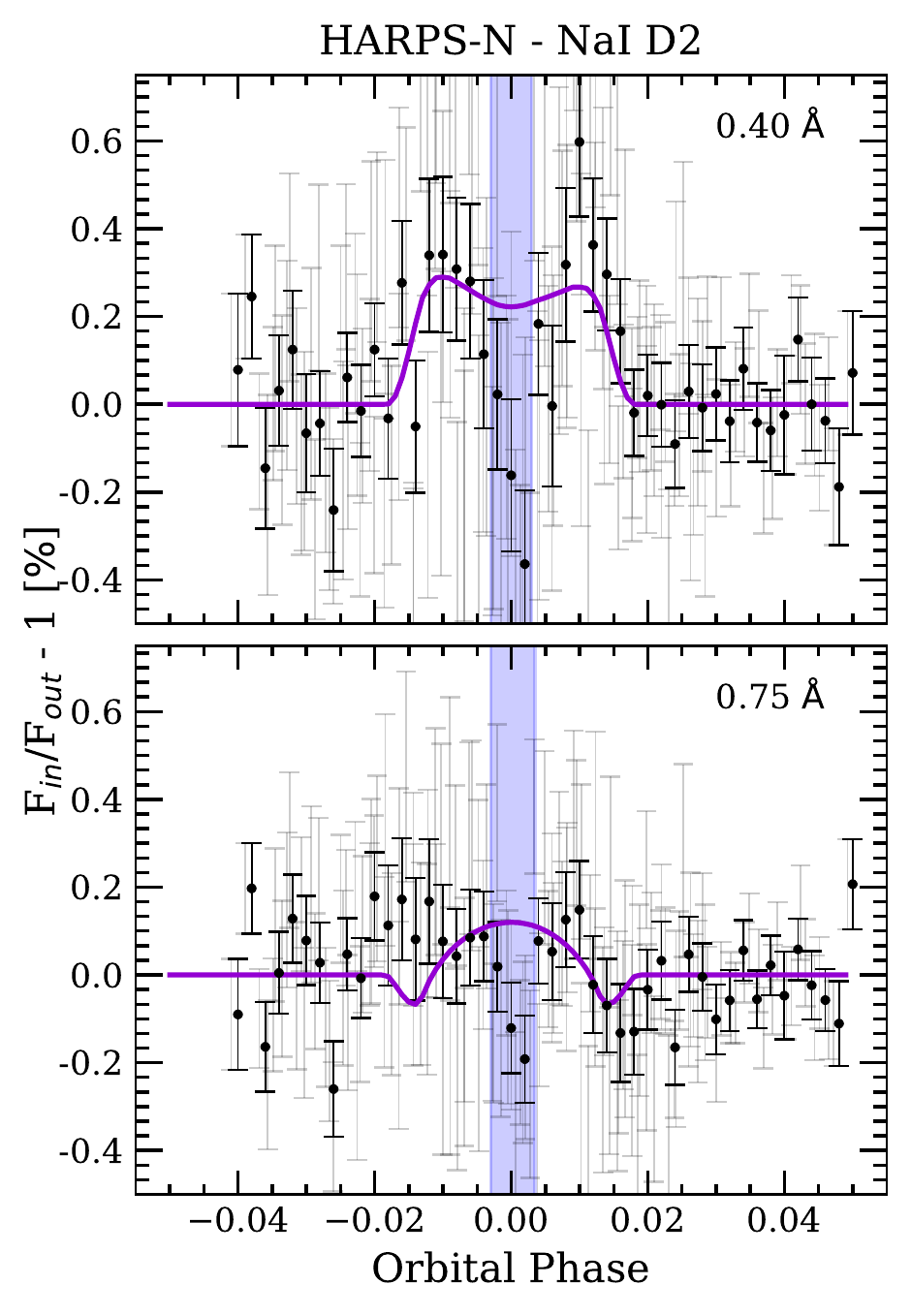}
\includegraphics[width=0.3\textwidth]{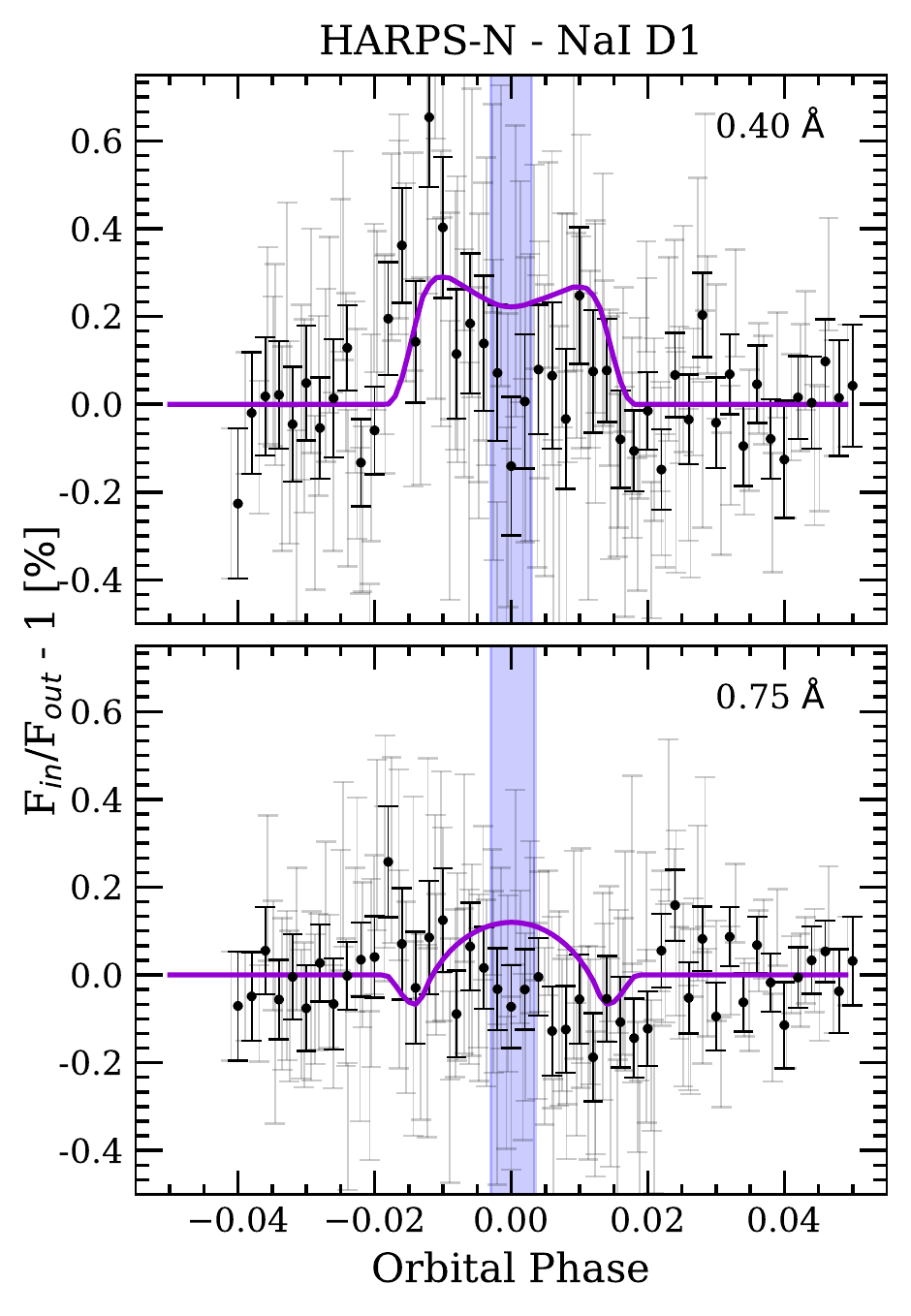}
\includegraphics[width=0.3\textwidth]{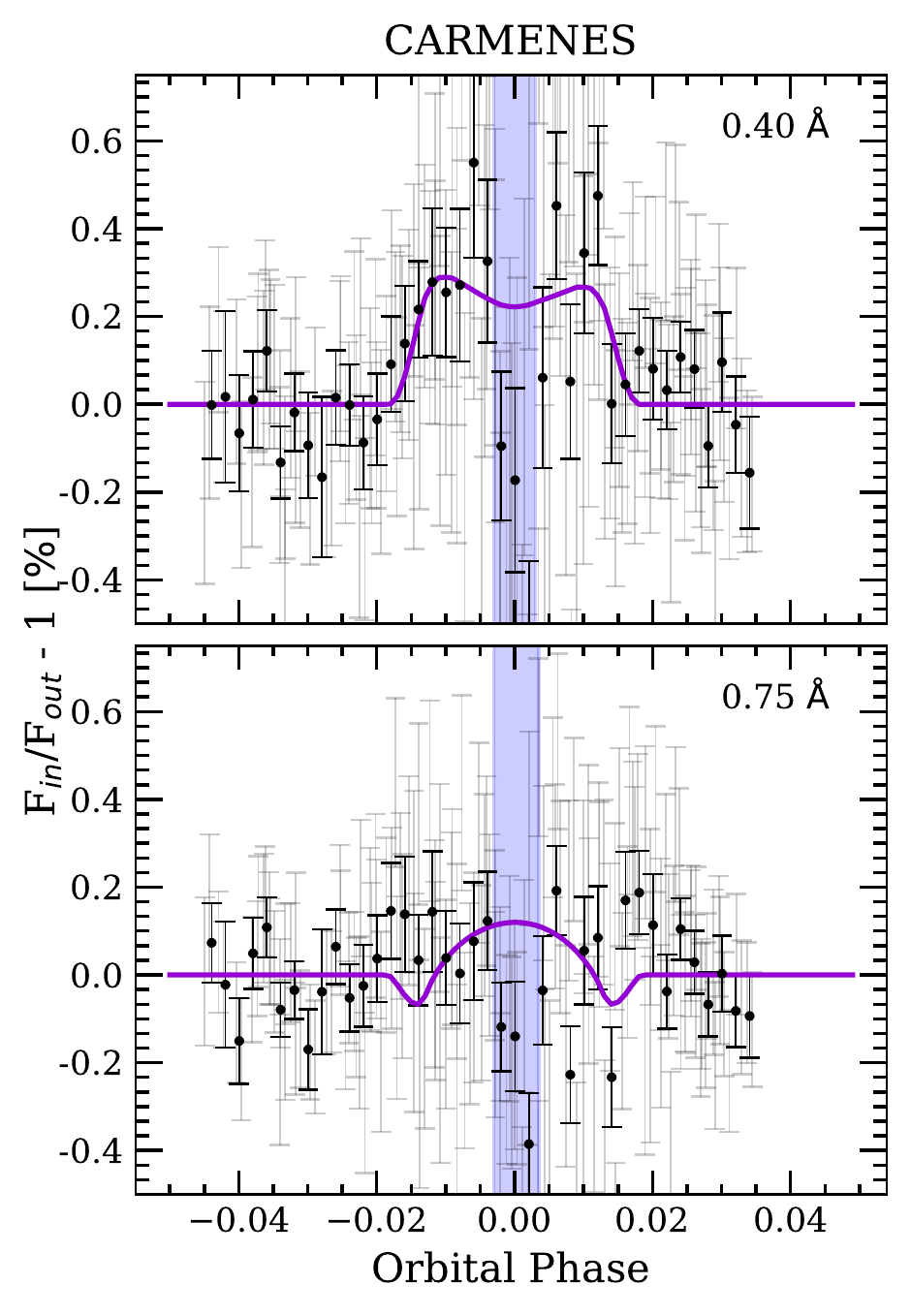}
\includegraphics[width=0.3\textwidth]{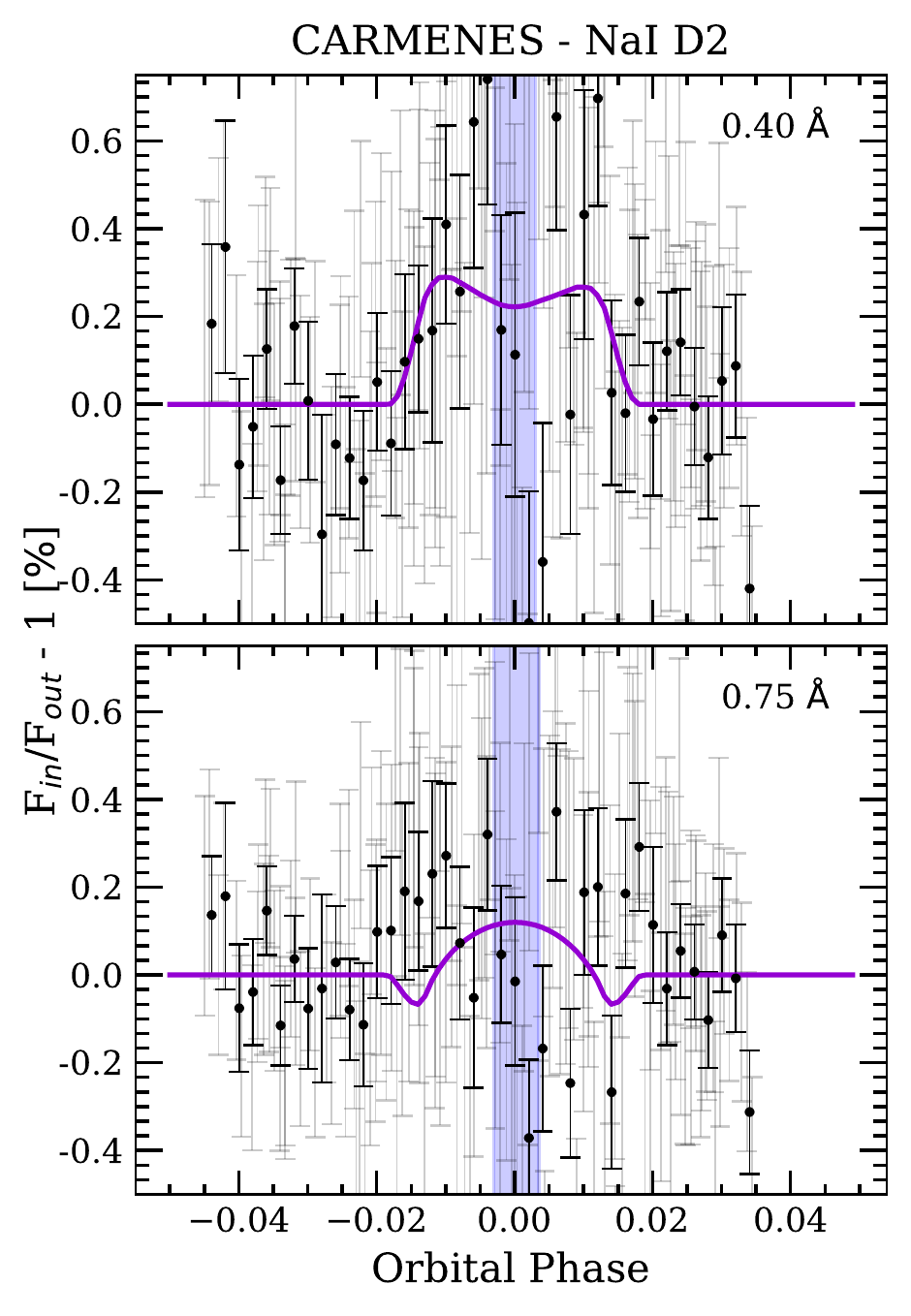}
\includegraphics[width=0.3\textwidth]{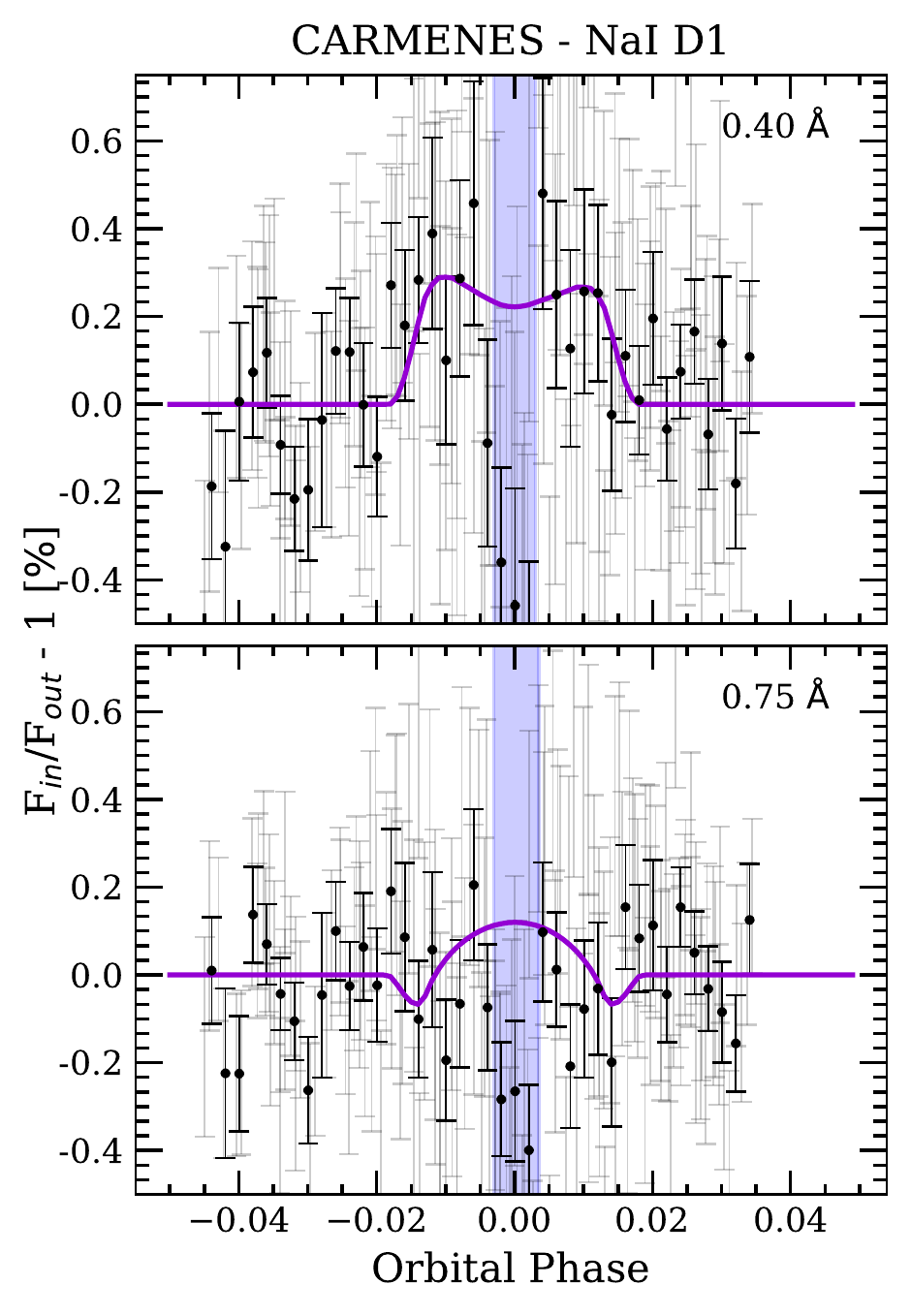}
\caption{Observed transmission light curves for the \ion{Na}{i} doublet computed using HARPS-N (top row panels) and CARMENES (bottom row panels) data sets. In each panel we show the light curves for two different passbands: $0.4~{\rm \AA}$ (\textit{top row}) and $0.75~{\rm \AA}$ (\textit{bottom row}). Each column corresponds to a different computation: the \ion{Na}{i} D2 and D1 lines combined (\textit{left column}), only \ion{Na}{i} D2 (\textit{middle column}), and only \ion{Na}{i} D1 (\textit{right column}). In all cases, the light grey data correspond to the original data, while the black dots are the data binned by $0.002$ in orbital phase. In purple we show the modelled light curves containing the CLV and RM effects.}
\label{fig:D21_TLC}
\end{figure}

\newpage

\section{Transmission light curves around \ion{Fe}{ii} $\lambda5169$, \ion{Mg}{i} $\lambda5173$, and \ion{Mg}{i} $\lambda5183$}
\label{ap:TLC_Fe}

\begin{figure}[h]
\centering
\includegraphics[width=0.3\textwidth]{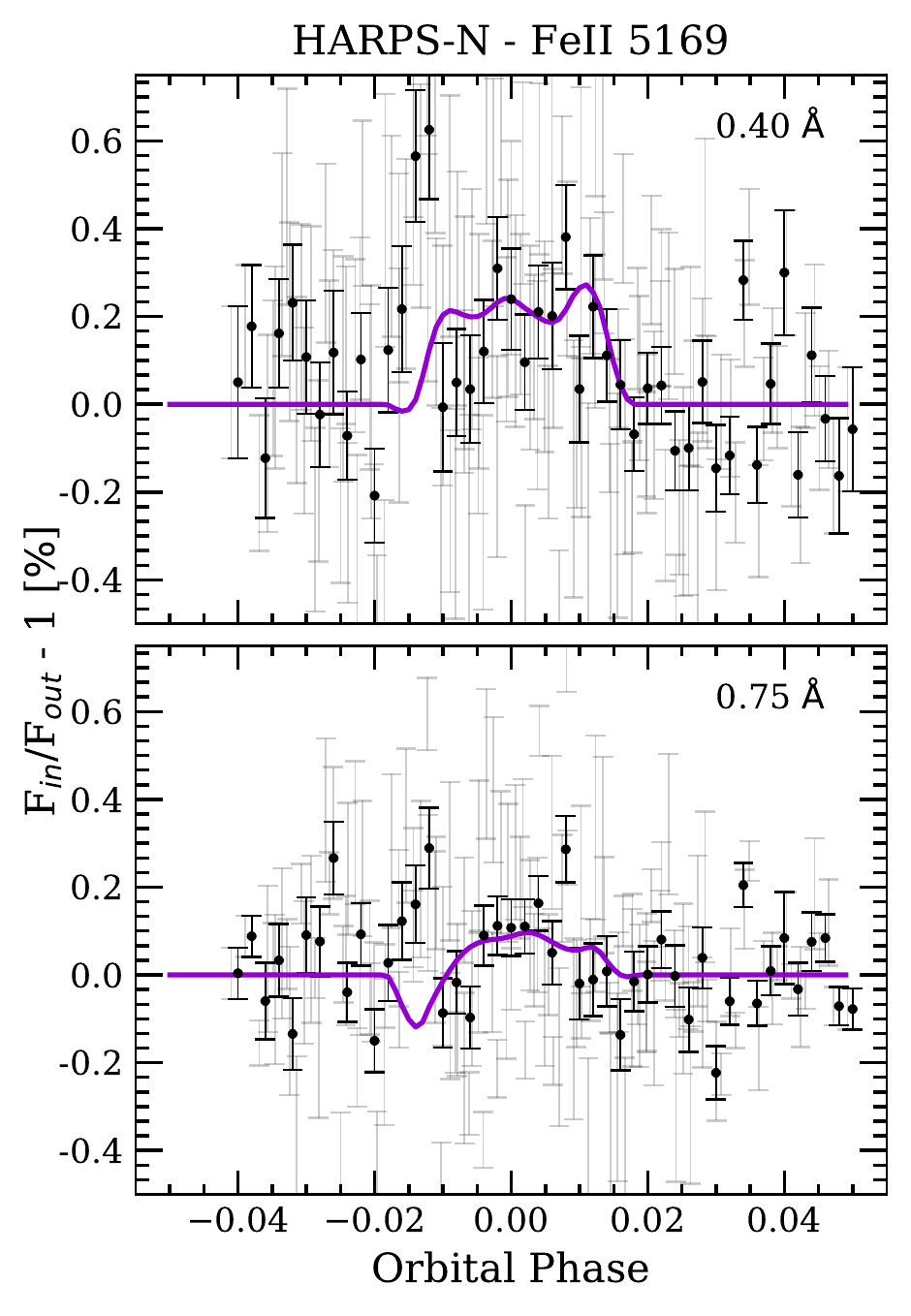}
\includegraphics[width=0.3\textwidth]{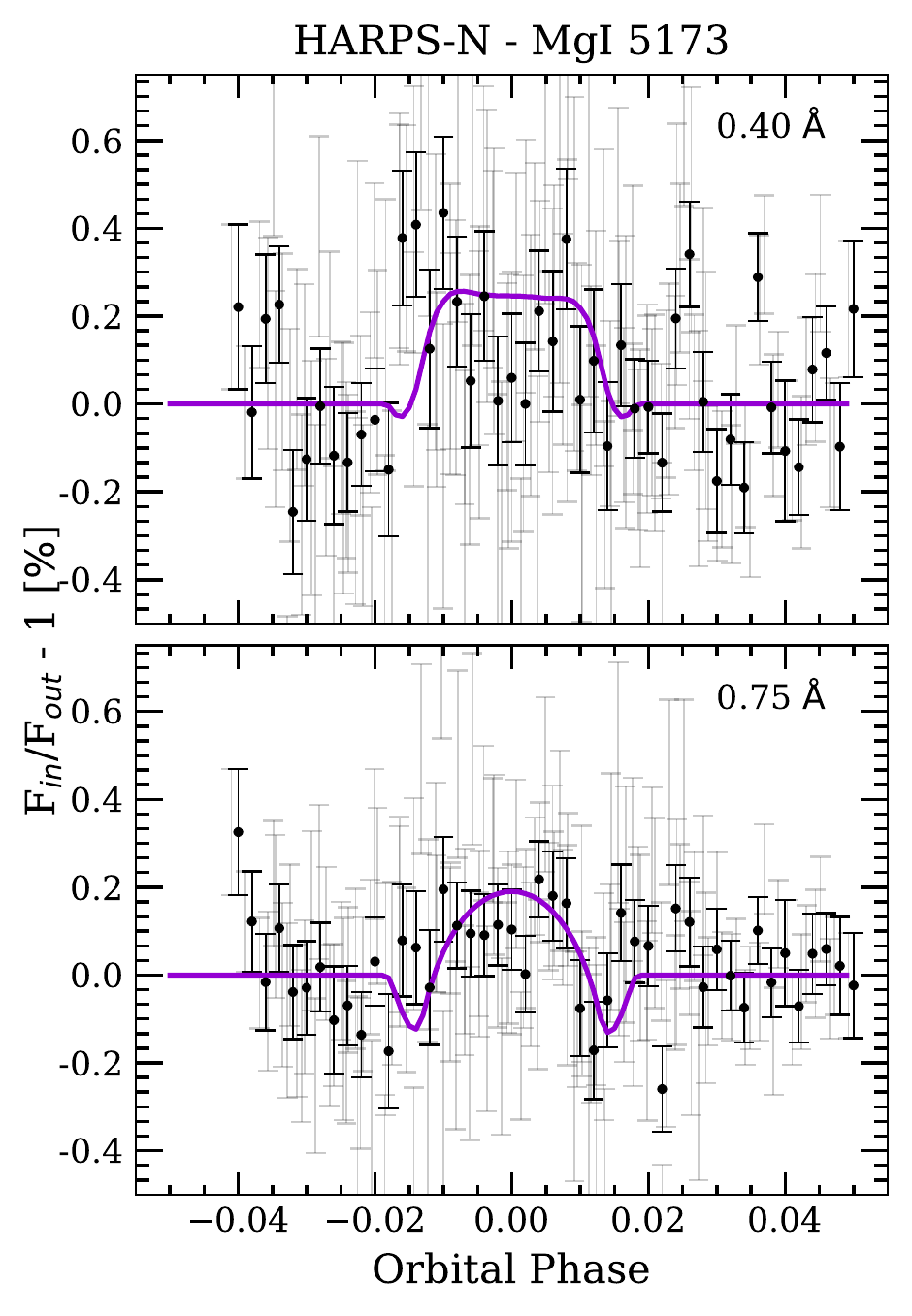}
\includegraphics[width=0.3\textwidth]{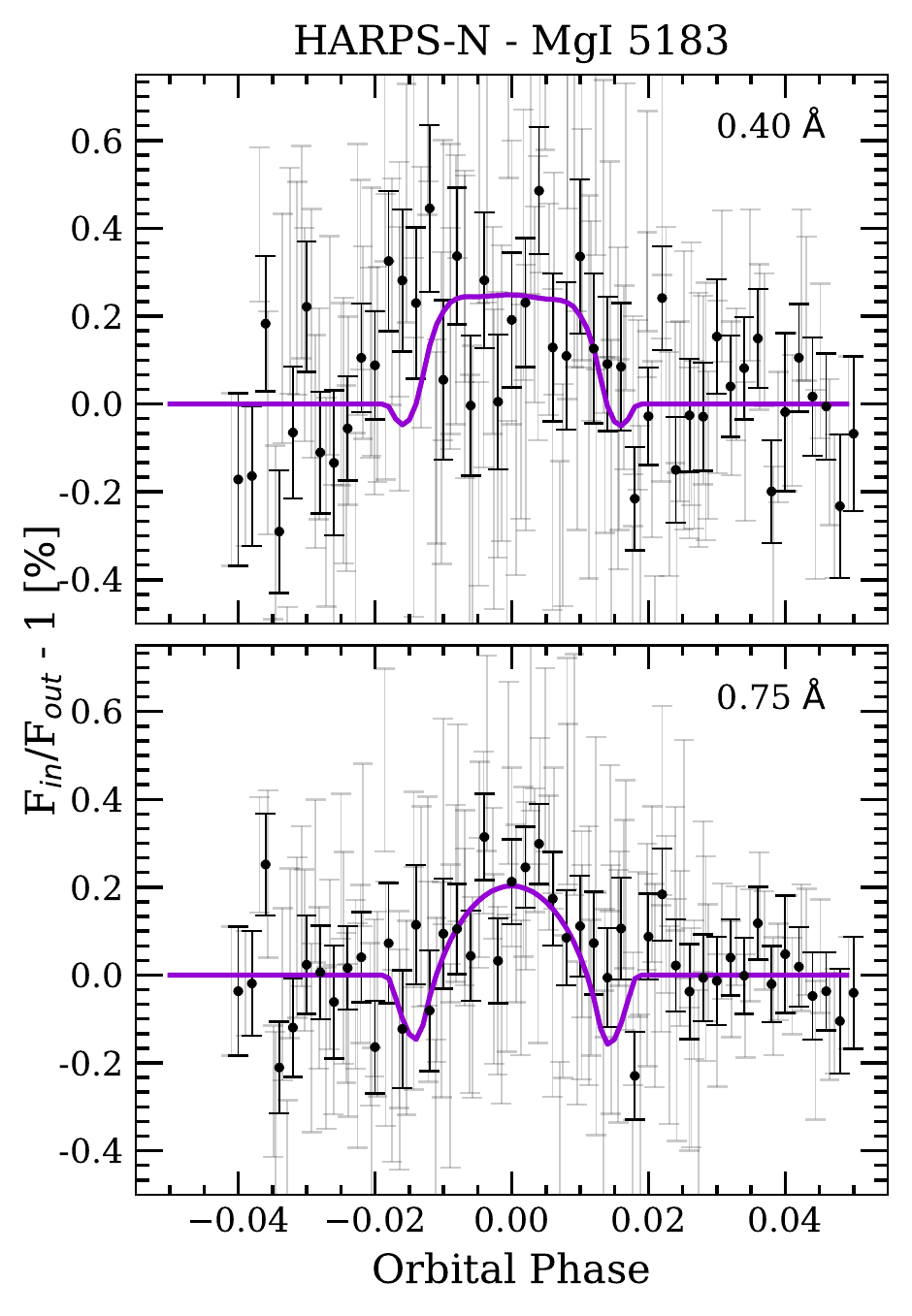}
\caption{Same as Figure~\ref{fig:D21_TLC}, but for the \ion{Fe}{ii} line at $5169~{\rm \AA}$ (left), \ion{the Mg}{i} line at $5173~{\rm \AA}$ (middle), and for \ion{Mg}{i} $5183~{\rm \AA}$ (right).}
\label{fig:Fe_TLC}
\end{figure}

\newpage

\section{Transmission spectra around H$\alpha$, \ion{K}{i} D1, and \ion{Ca}{ii} IRT}

\begin{figure}[H]
\centering
\includegraphics[width=0.7\textwidth]{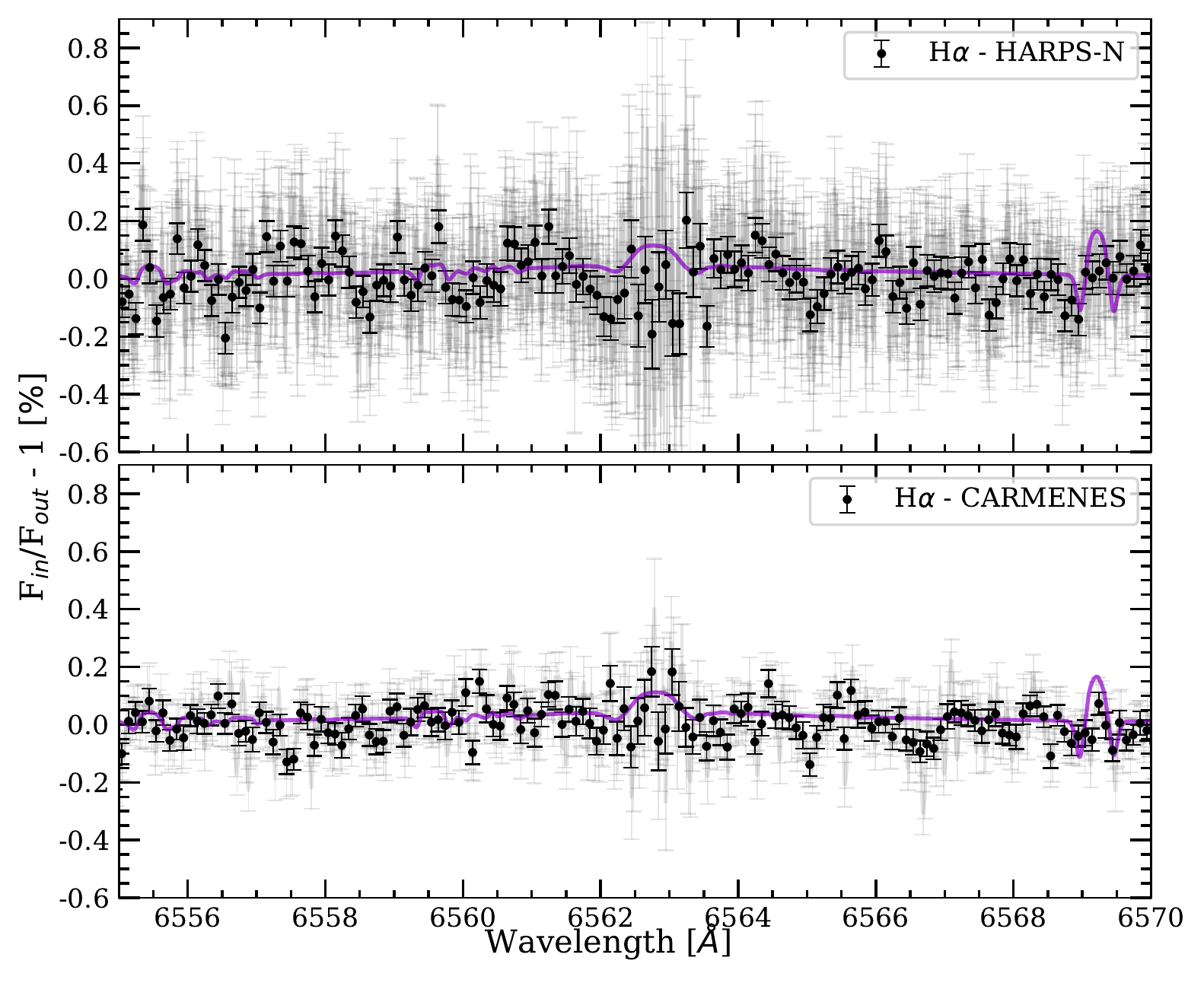}
\caption{Same as Figure~\ref{fig:indiv_TS_HARPS}, but around the H$\alpha$ line. In the top panel we show the results combining three HARPS-N transits, and in the bottom panel the results after combining two CARMENES observations.}
\label{fig:ts_Ha}
\end{figure}

\begin{figure}[H]
\centering
\includegraphics[width=0.7\textwidth]{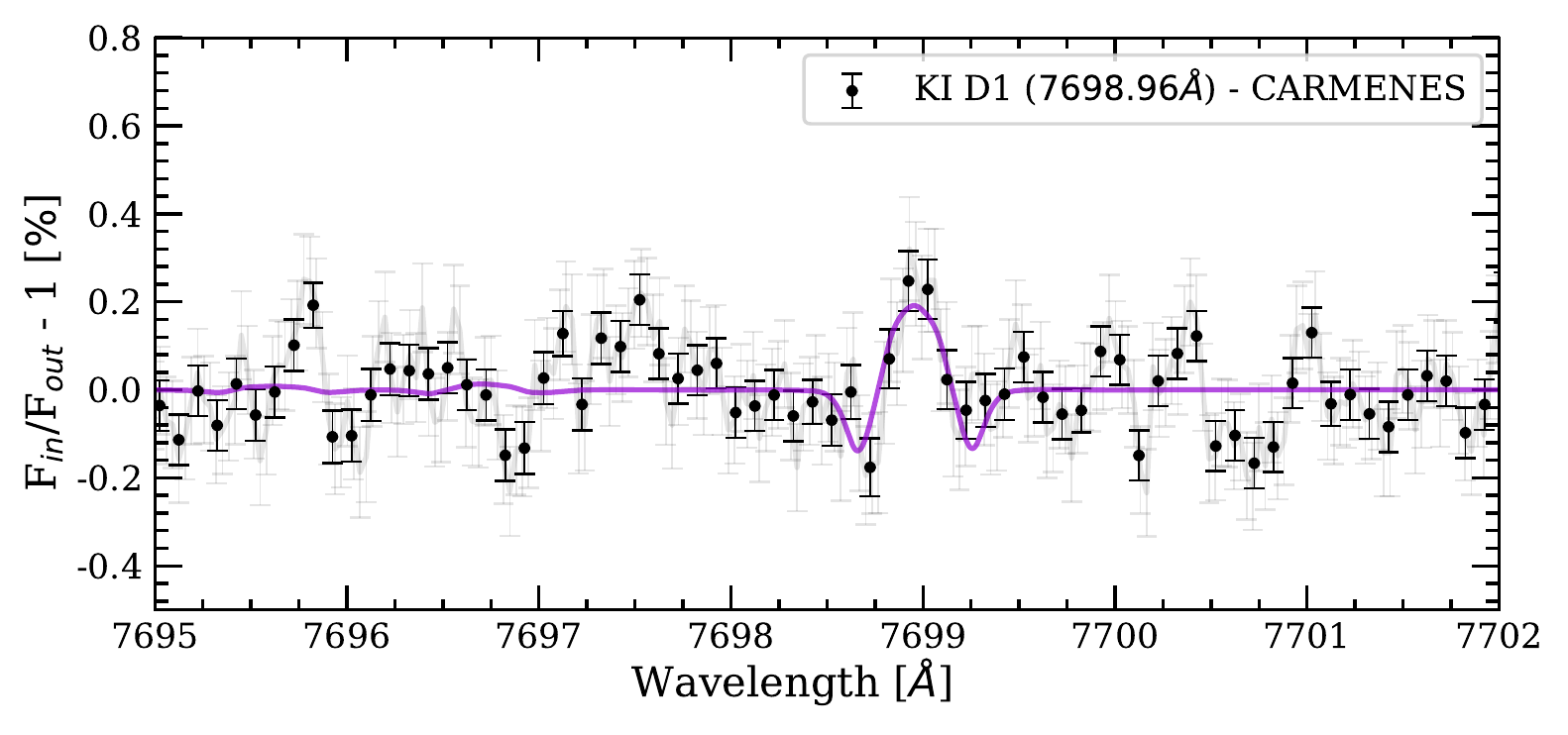}
\caption{Same as Figure~\ref{fig:indiv_TS_HARPS}, but around the \ion{K}{i} D1 line at $7698.96~{\rm \AA}$. This result is the combination of two CARMENES observations (HARPS-N does not cover this wavelength region).}
\label{fig:ts_K}
\end{figure}
\newpage

\begin{figure}[H]
\centering
\includegraphics[width=0.7\textwidth]{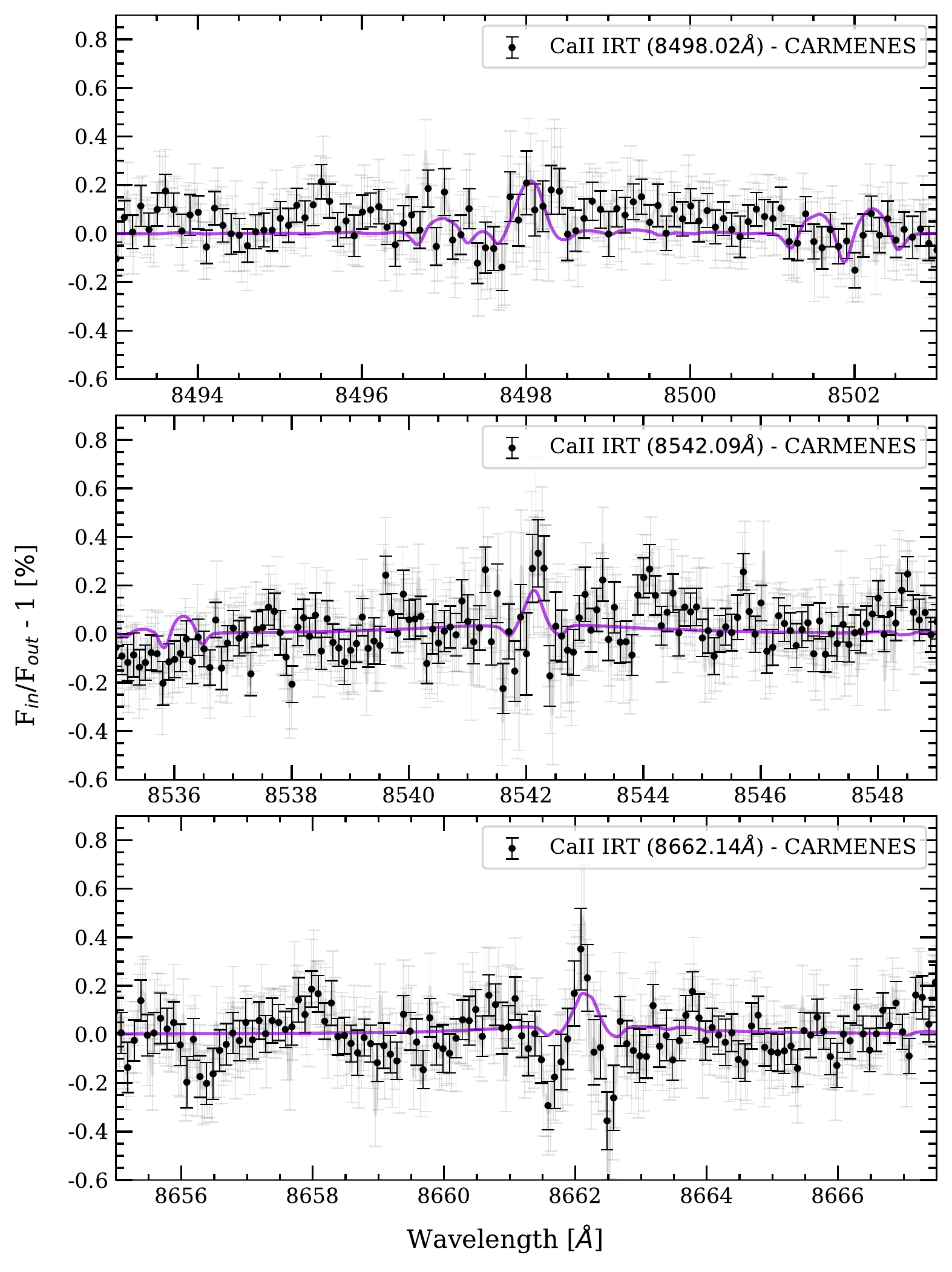}
\caption{Same as Figure~\ref{fig:indiv_TS_HARPS}, but around the \ion{Ca}{ii} IRT triplet. This result is the combination of two CARMENES observations (HARPS-N does not cover this wavelength region). In each panel we show the transmission spectrum around one of the \ion{Ca}{ii} IRT lines. Each line is specified inside the panel. We note that for these lines the model seems to be less intense than the data.}
\label{fig:ts_Ca}
\end{figure}

\end{appendix}

\end{document}